\documentclass[structabstract]{aa}  
\usepackage{graphicx}
\usepackage{txfonts}
\usepackage{multirow}
\usepackage{array}
\usepackage{natbib}
\begin{document}
   \title{Impact of angular differential imaging on circumstellar disk images}


   \author{J. Milli
          \inst{1}
          \and
		  D. Mouillet
		  \inst{1}
          \and
          A-M. Lagrange
		  \inst{1}
          \and
          A. Boccaletti\inst{2}
		  \and
		  D. Mawet
		  	\inst{3}
		  \and
		  G. Chauvin \inst{1}
		  \and
		  M. Bonnefoy \inst{4}
          }

   \institute{Institut de Planetologie et d'Astrophysique de Grenoble (IPAG), University Joseph Fourier,
              CNRS, BP 53, 
              38041 Grenoble, France\\
              \email{julien.milli@obs.ujf-grenoble.fr}
         \and
             LESIA, Observatoire de Paris, CNRS, Universite Pierre et Marie Curie 6 and Universite Denis Diderot Paris 7, 5 place Jules Janssen, 92195 Meudon, France
  		\and
  		European Southern Observatory, Casilla 19001, Santiago 19, Chile
  		\and
  		Max Planck Institute for Astronomy, Koenigstuhl 17, 69117 Heidelberg, Germany 
             }

   \date{Received 29 May, 2012; accepted 20 July, 2012}

 
  \abstract
   {Direct imaging of circumstellar disks requires high-contrast and high-resolution techniques. The angular differential imaging (hereafter ADI) technique is one of them, initially developed for point-like sources but now increasingly applied to extended objects such as disks. This new field of application raises many questions because the disk images reduced with ADI depend strongly on the amplitude of field rotation and the ADI data reduction strategy. Both of them  directly affect the disk observable properties.
   }
   {Our aim is to characterize the applicability and biases of some ADI data reduction strategies for different disk morphologies. A particular emphasis is placed on parameters mostly used for disks such as their surface brightness distribution, their width if the disk is a ring, and local features such as gaps or asymmetries. We first present a general method for predicting and quantifying those biases. In a second step we illustrate them for some widely used ADI algorithms applied to typical debris disk morphologies: inclined rings with various inner/outer slopes and width. Last, our aim is also to propose improvements of classical ADI to limit the biases on extended objects.
   }
   {Simulated fake disks seen under various observing conditions
    were used to reduce ADI data and quantify the resulting biases. These conclusions are complemented by previous results from NaCo L' real-disk images of HR\,4796A.	
   }
   {As expected, ADI induces flux losses on disks. This makes this technique appropriate only for low- to medium-inclination disks. A theoretical criterion is derived to predict the amount of flux loss for a given disk morphology, and quantitative estimates of the biases are given in some specific configurations. These biases alter the disk observable properties, such as the slopes of the disk surface brightness or the radial/azimuthal extent of the disk. Additionally, this work demonstrates that ADI can very easily create artificial features without involving astrophysical processes. For example, a particularly striking feature appears for a ring when the amplitude of field rotation is too small. The two ring ansae are surrounded by two flux-depleted regions, which makes them appear as bright blobs. This observation does not require any astrophysical process such as dust blown by radiation pressure, as proposed previously in H-band images of HR\,4796A.  
   }
  {The ADI techniques behave as spatial filtering algorithms and can bias disk observables. Therefore, the filtering process needs to be properly calibrated when deriving disk parameters from processed images. 
  }
  

   \maketitle
%

\section{Introduction}

The study of circumstellar disks is essential for understanding the formation of planetary systems. Direct imaging of these disks has revealed asymmetries, warps, gaps, troncatures, density waves, or other features that are the results of interactions between the disk and its environment. About 160 disks are now resolved from their visible to thermal emission (http://www.circumstellardisks.org, Stapelfeldt). Debris disks are a particularly interesting class of optically thin disks since planets are already formed, if any, and faint structures in the dust distribution can betray their presence, as proposed for HR\,4796A in \citet{Lagrange2012b}.  

The vicinity of the star and its luminosity however limit the performance of both ground-based \citep{Racine1999} and space-based \citep{Schneider2003} high-contrast imaging because of bright quasi-static speckles. Slowly evolving, they add up over time and eventually become the dominant noise source at separations below a few arcseconds, depending on the star brightness \citep{Macintosh2005}
  
The principle of differential imaging is to subtract a reference frame from the target image to reduce quasi-static speckle noise \citep{Marois2006}. This reference frame is sometimes also called a reference Point Spread Function \citep{Lafreniere2007}, hereafter PSF, but we will use here the more generic term reference frame. This frame can be obtained in various ways: either with a reference star or with the target itself observed at a different field of view orientation (ADI), a different wavelength (spectral differential imaging), or polarization (polarimetric differential imaging). This paper focuses on ADI. 

Angular differential imaging has already achieved significant results on debris disks: \citet{Buenzli2010} for instance detailed the morphology of the Moth, \citet{Boccaletti2012} and \citet{Currie2012} revealed the inner part of HD\,32297, \citet{Thalmann2011} and \citet{Lagrange2012b} studied the ring of HR\,4796A,  \citet{Lagrange2012} revealed the inner part of the $\beta$ Pictoris disk and constrained the projected planet position relative to the disk, and \citet{Thalmann2010} resolved the gap in the transitional disk around LkCa\,15.
However, side effects might alter the apparent morphology of the disk and its observable properties: flux self-subtraction, change in disks slopes, width, etc. The model-matching procedure described by \citet{Boccaletti2012} to capture the true morphology of HD\,32297 is a good illustration of how difficult it is to dismiss ADI artifacts. 

Angular differential imaging uses the differential rotation between the field of view and the pupil occurring on an alt-azimuthal mount to distinguish between the speckle halo and any on-sky source. When the pupil tracking mode is used, the field rotates at the same rate as the parallactic angle while the pupil is stable. The parallactic angle is the angle between the great circle through the object and the zenith, and the hour circle of the object.

When extended objects such as disks are imaged, the challenge of ADI is to build reference frames with a speckle pattern highly correlated to the target image but without capturing the flux of the disk. If the disk flux is captured in the reference frame, the reference subtraction will decrease the disk flux in the residual image: this is the problem of self-subtraction. Not only does it lead to a lower signal-to-noise ratio (S/N) for the disk but it also biases the observable parameters of the disk.

The objective of this study is twofold: 
\begin{enumerate}
\item To guide the observing strategy and data reduction by providing key figures to know how much flux loss is expected for a given disk geometry and field of view rotation. A theoretical criterion is derived to quantify this loss for the specific case of an edge-on disk.
\item To highlight the observable parameters that can be trusted from ADI-based disk images and point out the artifacts potentially created by this technique.
\end{enumerate}

We will first describe the simulation procedure, and in Section 3 will build theoretical criteria to analyze the applicability of ADI to disks. A qualitative (Section \ref{SectionQualitative}) and quantitative (Section \ref{SectionQuantitative}) description of the biases induced by ADI is then presented before reviewing specific ADI algorithms more adapted for disk reductions in Section \ref{SectionImprovements}. Systematically investigating  the parameter space of these algorithms to derive the most accurate disk parameters is beyond the scope of that section.

\section{ADI algorithms and simulation procedure}

Several data reduction strategies to treat ADI data have been developed over the past few years. They differ in the way the images are selected and combined to build a reference frame. The selection can use all available images or only a reduced sample based on a disk binary mask, depending on how much the parallactic angle has changed from the working frame, how close in time the image is from the working frame, or how correlated they are. The combination can be a simple median, an average, or a linear combination of all selected frames.   

We focus here on three specific algorithms referred to as classic ADI (cADI), radial ADI (rADI) and LOCI. These procedures are described in \citet{Lagrange2012}. We will review three additional techniques specific for disks that use iterations, binary masks for the disk (mcADI, mLOCI), and damped LOCI (dLOCI).

\begin{itemize}
\item In cADI \citep{Marois2006}, the reference frame is a simple median or mean of all individual images of the cube. The key parameter is the total amount  of field rotation available, $\Delta \theta$, typically $20^\circ$ to $80^\circ$ for observations of one to several hours.
\item In rADI, the reference frame is estimated radially in annuli as presented in \citet{Marois2006}. The key parameters are the separation criterion $N_\delta$, usually 1 to 5 FWHM, the number of images $N$ used to build one reference frame (between 5 and 30) and the radial extent of the annulus $dr$. For each working frame, the reference frame is obtained by taking the mean of the selected $N$ images. The selection of the N images is made temporally, after sorting the images according to their hour angle and selecting the $N$ closest to the current working frame for which the field has rotated more than the separation criterion $N_\delta$. 
\item In LOCI \citep{Lafreniere2007}, the reference frame is estimated locally in subtraction areas using linear combinations of all sufficiently rotated frames. The linear coefficients, called $c^k$, are computed to minimize the residuals in an optimization area. The geometry of the optimization area is an annulus section parametrized through the aspect ratio g, and the surface expressed in number of PSF cores $N_A$. The geometry of the subtraction area is parametrized through the annuli width $dr$. One key parameter is the separation criterion $N_\delta$ as in rADI. 
 \item McADI and mLOCI are variants of cADI and LOCI using masks to limit the contamination of the reference frame by the disk. They are detailed in Section \ref{SubsectionMasking}.
 \item DLOCI is an algorithm detailed in \citet{Pueyo2012}. It is applied here to disks and is discussed in Section \ref{SubSectiondLOCI}
\end{itemize}

To study the artifacts generated by those algorithms, we generated fake disk images, inserted them into data cubes at the appropriate parallactic angle and convolved them with a synthetic or real telescope PSF. In some cases, noise coming from real data was also added to these simulation data cubes.  
The fake disks were assumed to be axisymmetric and were generated from an initial dust volumetric density $\rho(r,z)=\rho_0 R(r) Z(r,z)$ with the simulation code GRaTer \citep{Augereau1999}. $R(r)=\left\lbrace \left( \frac{r}{r_0} \right)^{-2\alpha_{in}} + \left( \frac{r}{r_0} \right) ^ {-2\alpha_{out}} \right\rbrace ^{-0.5}$ is the radial dust distribution in the mid-plane and $Z(r,z)$ is the vertical dust distribution at radius $r$. 


\section{Applicability of ADI to disks: the problem of self-subtraction}

When the disk azimuthal extent is large with respect to the field rotation, the reference frame used to subtract the light of the central star and reveal circumstellar matter contains some flux from the disk and self-subtraction occurs.

\subsection{Pole-on disks}
An axisymmetric disk seen pole-on is entirely self-subtracted, because the surface brightness distribution (hereafter SBD) is symmetrical with respect to the central star. 
In this configuration, ADI is unsuitable and the PSF subtraction should be made using a reference star observed close to the target, preferably at the same parallactic angle, to maximize speckle correlation \citep{Beuzit1997}. Nevertheless, non-axisymmetric features such as bright blobs can still be revealed through ADI, but this application remains limited. 

\subsection{Edge-on disks}
\begin{figure}
   \centering
   \includegraphics[width=6cm]{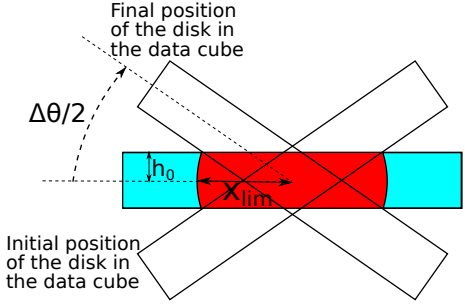}
      \caption{Uniform disk of  height $2h_0$ seen edge-on at the beginning, middle and end of the observing sequence. In red, the final disk image after cADI processing, in blue the part of the disk that is entirely self-subtracted.}
         \label{FigUniformDisk}
   \end{figure}

The situation is different for disks seen edge-on such as the $\beta$ Pictoris or AU Mic disk. The key parameter is the vertical extent of the disk at a separation $x$ relative to the field rotation at that radius. In the very theoretical case of a uniform disk of vertical scale height $h_0$, with a total field rotation $\Delta \theta$, and a reference frame built by median-combining all available images (cADI), the residual disk is entirely self-subtracted for a separation $x<x_{lim}=\frac{h_0}{tan(\Delta \theta /4)}$, as illustrated in Figure \ref{FigUniformDisk}.

\begin{figure}
	\centering
	\includegraphics[width=\hsize]{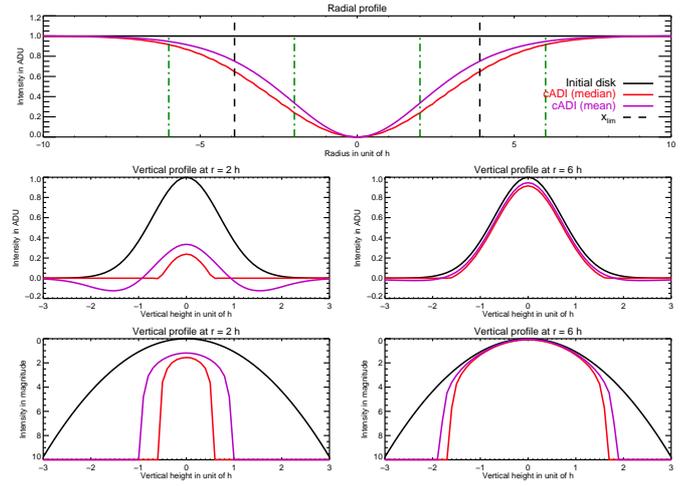}
	\caption{Top: radial intensity profile along the disk mid-plane of the initial disk (black) and cADI images (reference frame built by median-combination in red and by mean-combination in purple). Middle: two vertical intensity profiles at the radii corresponding to the green dashed lines of the top figure, in linear scale. Bottom: same vertical profiles in magnitude scale.}
	\label{FigCADIRegimes}
\end{figure}

For a non-uniform disk, with a vertical profile parametrized with an exponential index $\gamma$ so that $I(z)=e^{-{\left( \frac{z}{h} \right)} ^\gamma}$, there is no longer a sharp boundary radius $x_{lim}$ between a totally preserved region and a totally self-subtracted region. However, we can identify three different regimes, depending of the scale height $h$ at the given radius with respect to $\Delta \theta$. These are illustrated in Figure \ref{FigCADIRegimes} for an initial disk of vertical scale height $h$, with an exponential index $\gamma=2$ (Gaussian vertical profile) and a field rotation $\Delta \theta=60^\circ$. Either the disk is well preserved in its whole vertical extent (for a radius beyond \textbf{$7 \times h$}) , or the vertical extent is reduced while the mid-plane brightness is well preserved (between 7 and $6 \times h$) or the mid-plane is also dimmer (below \textbf{$6 \times h$}). The last two regimes are illustrated by vertical profiles shown in linear and magnitude scales in the last four graphs of Figure \ref{FigCADIRegimes}. These graphs show that brightness profiles can vary significantly whether the reference is combined by median or mean. With a mean, the flux in the mid-plane is higher (top chart of left graph of Figure \ref{FigCADIRegimes}) and the vertical extent of the disk is also larger but negative regions appear on each side of the mid-plane, as visible in linear scale in the middle left graph of Figure \ref{FigCADIRegimes}. A median prevents negative regions from appearing on noiseless data, or limits their extension in the presence of noise and is more robust to noise when many frames are available for reference frame combination. Unless otherwise specified, the reference frame is built by median combination in cADI throughout. 

A theoretical criterion to quantify the disk flux loss can be derived in this specific vertical brightness distribution. Let $\Delta\alpha = arctan \left( \frac{h}{r} \right)$ be the angular vertical height of the disk seen from the star. We aim to characterize the remaining disk flux at the point of vertical height z and radius $r$, corresponding to an angular height $i = arctan \left( \frac{z}{x} \right)$. 
The  remaining disk flux $f$ at a given position in the reduced cADI image, normalized to the initial disk flux (for a reference frame built by  median-combination), follows the exponential law,
\begin{equation}
f(\kappa_c) = 1- e^{ -\kappa_c} \mbox{ with }  \kappa_c(i) = \frac{ (\Delta \theta/4)^\gamma -i^\gamma}{(\Delta \alpha)^\gamma}. 
\label{EquFluxLoss} 
\end{equation}
For instance, in the disk mid-plane ($i=0 ^{\circ}$), $\kappa_c = \left( \frac{ (\Delta \theta/4)}{\Delta \alpha} \right)^\gamma $ and for $\kappa_c = 3$, 95\% of the disk flux is preserved. The derivation of that law is detailed in Appendix \ref{AppendixFluxLossCriteria} and is verified experimentally. 

A similar criterion can be derived for an edge-on disk with exponential vertical profile processed with rADI. In this case, however, the field rotation term $\Delta \theta$ is replaced by a term depending on the separation criterion  $N_{\delta}$ and the  number of images $N$ used to build the reference PSF. If we call $\Delta \epsilon = \frac{\Delta \theta}{N_{total}} \times N $
the mean field rotation corresponding to N images and $\sigma = arctan \left( \frac{N_{\delta}\times {FWHM}}{r} \right) $ the separation criterion expressed in terms of angle, the disk flux $f$ in the mid-plane follows the law

\begin{equation}
f(\kappa_r) = 1- e^{ -\kappa_r} \mbox{ with }  \kappa_r = \left(\frac{\sigma + \Delta \epsilon/2}{\Delta \alpha} \right)^\gamma. 
\end{equation}

\subsection{General case}
\label{SubsectionTheoreticalModel}

In the two previous subsections, we showed that ADI is inappropriate for pole-on disks and we derived the key criterion $\kappa$ that allows one to predict the amount of flux losses for an axisymmetric edge-on disk.

In the general case of a non-axisymmetric disk with a random inclination $i$, the amount of flux loss cannot be summarized by this straight-forward mathematical criterion. Instead, we show below that ADI can be seen as a spatial frequency filtering algorithm, and we use this study to answer the question to what extent ADI can be applied.

\paragraph{ADI, a spatial frequency filtering algorithm:}

To show that cADI can be seen as a filtering algorithm, we built a frequency filter based on Fourier analysis and applied it to disk models to compare the results with cADI-processed images. 
The tool uses as input a simulated disk image and an amplitude of field rotation $\Delta \theta$ and performs the following steps for each pixel of coordinate $(r,\theta)$ in the image: 
\begin{itemize}
\item it extracts the 1D-azimuthal profile at the radius $r$
\item it sets to zero the points of the profile whose distance to $\theta$ is greater than $\Delta\theta / 2$ 
\item it builds the Fourier spectrum of the resulting profile and removes from this spectrum any frequencies lower than a predefined cutoff frequency $u_0$ 
\item last, it reconstructs the azimuthal intensity profile by an inverse Fourier transform of the spectrum, and reads the intensity at the pixel position $(r,\theta)$.
\end{itemize}
Different values for the cutoff frequency $u_0$ were tested and the choice of $\frac{1}{\Delta \theta}$ leads to an image close to the cADI processed image, when the reference frame is estimated by mean combination. This means that cADI can be seen as a low-pass filter of the azimuthal frequencies of an image, the cutoff frequency being that corresponding to the total amplitude of field rotation $\Delta \theta$.  

\begin{figure}
   \centering
    \includegraphics[width=9cm]{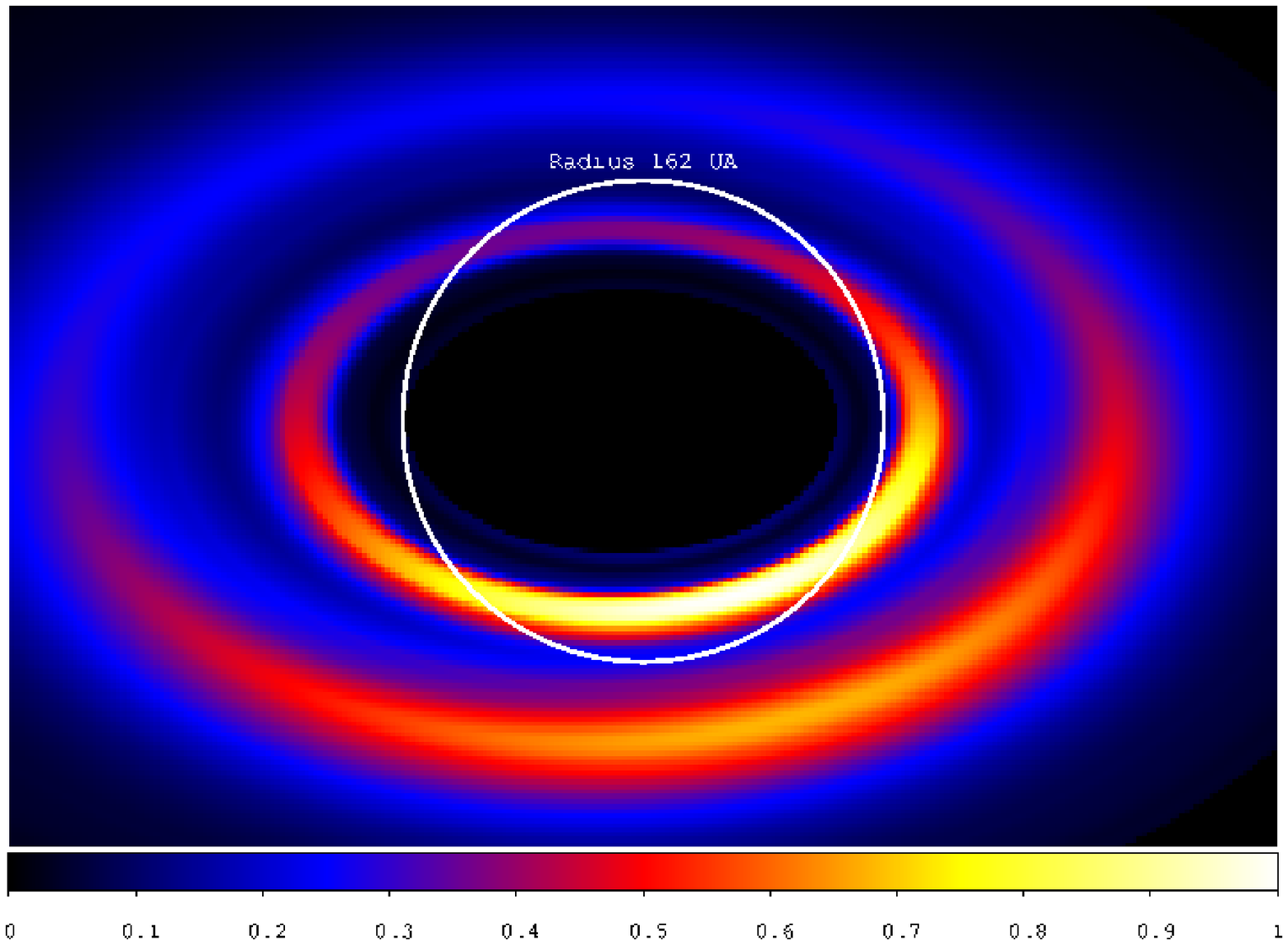}    
   \includegraphics[width=9cm]{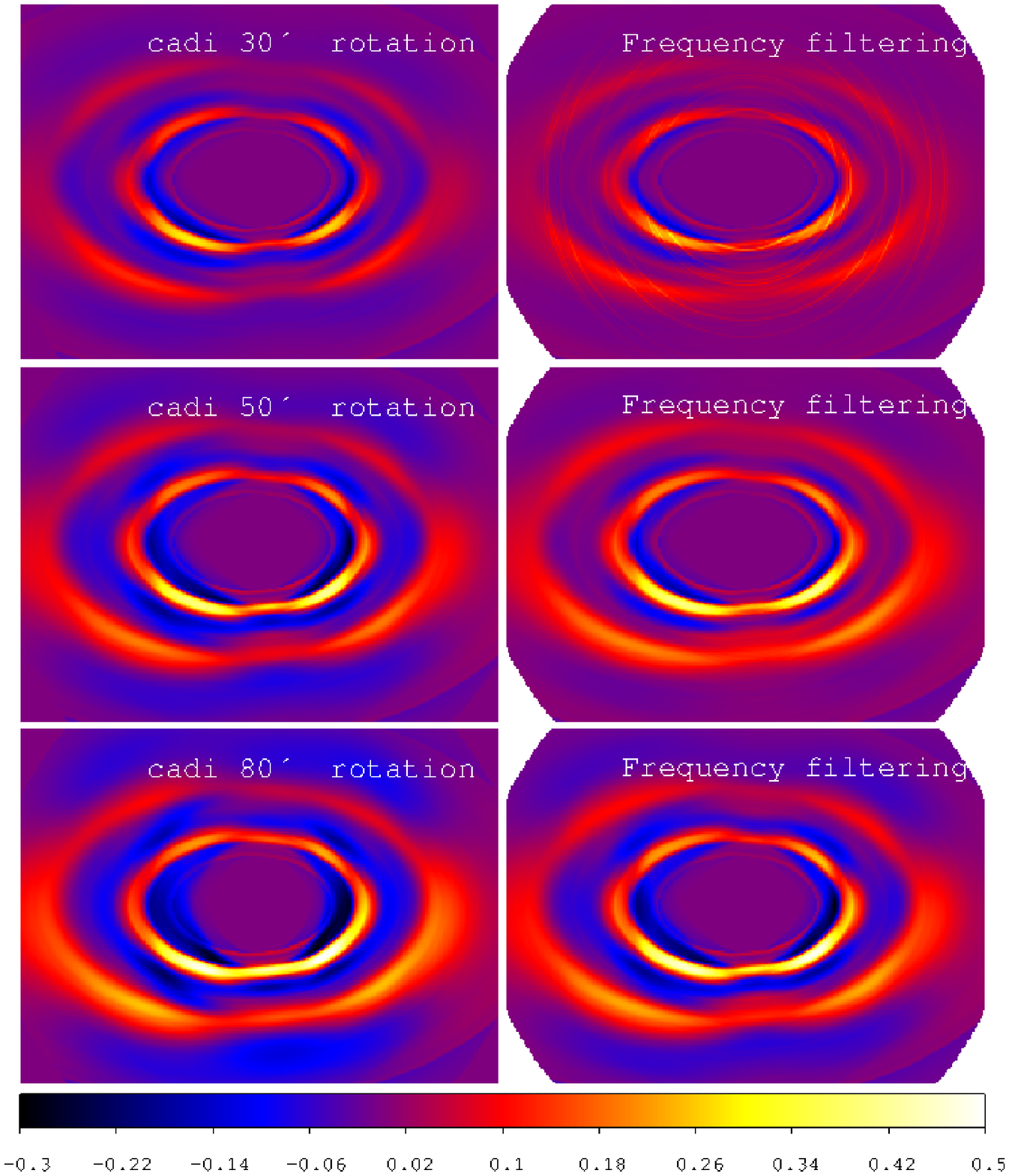}
   \caption{HD\,141569 disk model (top), and a comparison between cADI processed images with different field rotations and residual flux as predicted by the frequency filter. The color scale is identical and linear for all images.}
         \label{FigHD141569CADIFTpredicted}
   \end{figure}

A comparison between disk images processed by cADI and by our frequency-filtering algorithm is given in Figure \ref{FigHD141569CADIFTpredicted} for a theoretical model of the debris disk HD\,141569. We used three different amplitudes of field rotations: $30^\circ$, $50^\circ$ and $80^\circ$.

   \begin{figure}
   \centering
   \includegraphics[width=9cm]{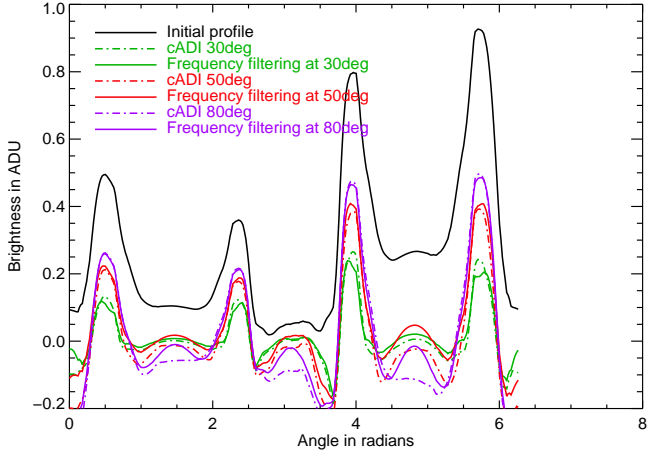}
   \caption{Comparison between azimuthal brightness profiles of HD141569 at a radius of 162 AU, initially and after various cADI processing or frequency filtering. 0 rad corresponds to the right panel in Figure \ref{FigHD141569CADIFTpredicted}.}
   \label{FigHD141569profiler38}
   \end{figure}

The disk inclination is $52^\circ$, which is the critical value below which ADI induces so much self-subtraction to make it inappropriate. The scattered light model of that disk was made using the GRaTer code, starting from a radial distribution for the disk optical depth, using a disk ellipticity of 0.1, an anisotropic scattering factor of -0.2, a Gaussian vertical profile,  and a constant opening angle. The initial model is displayed at the top of Figure \ref{FigHD141569CADIFTpredicted}. The disk contains several rings and anisotropic scattering makes the side inclined toward the line of sight brighter. The image of the disk processed by cADI shows that the morphology of the two rings has been altered. They are dimmer and the brighter region of those two rings is thinner, distorted and not any longer the brightest region. Increasing the total amount of field rotation yields more disk flux but does not prevent the distortions created in the disks. 	
All these observations are also valid for the images created by the filtering algorithm. Therefore we can conclude that they result from filtering the low spatial frequencies that have an orthoradial component.
Areas with negative flux between the two rings also appear. These negative areas are better visible in the brightness azimuthal profile at $r=162 AU$ shown in dashed lines in the plot of Figure \ref{FigHD141569profiler38}. They are well reproduced by the filtering algorithm (plain lines in Figure \ref{FigHD141569profiler38}), confirming the validity of the interpretation of cADI in terms of frequency filter.

\section{Qualitative and morphological effects of ADI on disks}
\label{SectionQualitative}

\subsection{Negative flux artifacts: extinctions, blobs and asymmetries}

Most of the artifacts created in ADI are due to over-subtracted areas that can make fake structures appear.

\begin{figure}
   \centering
       \includegraphics[width=8cm]{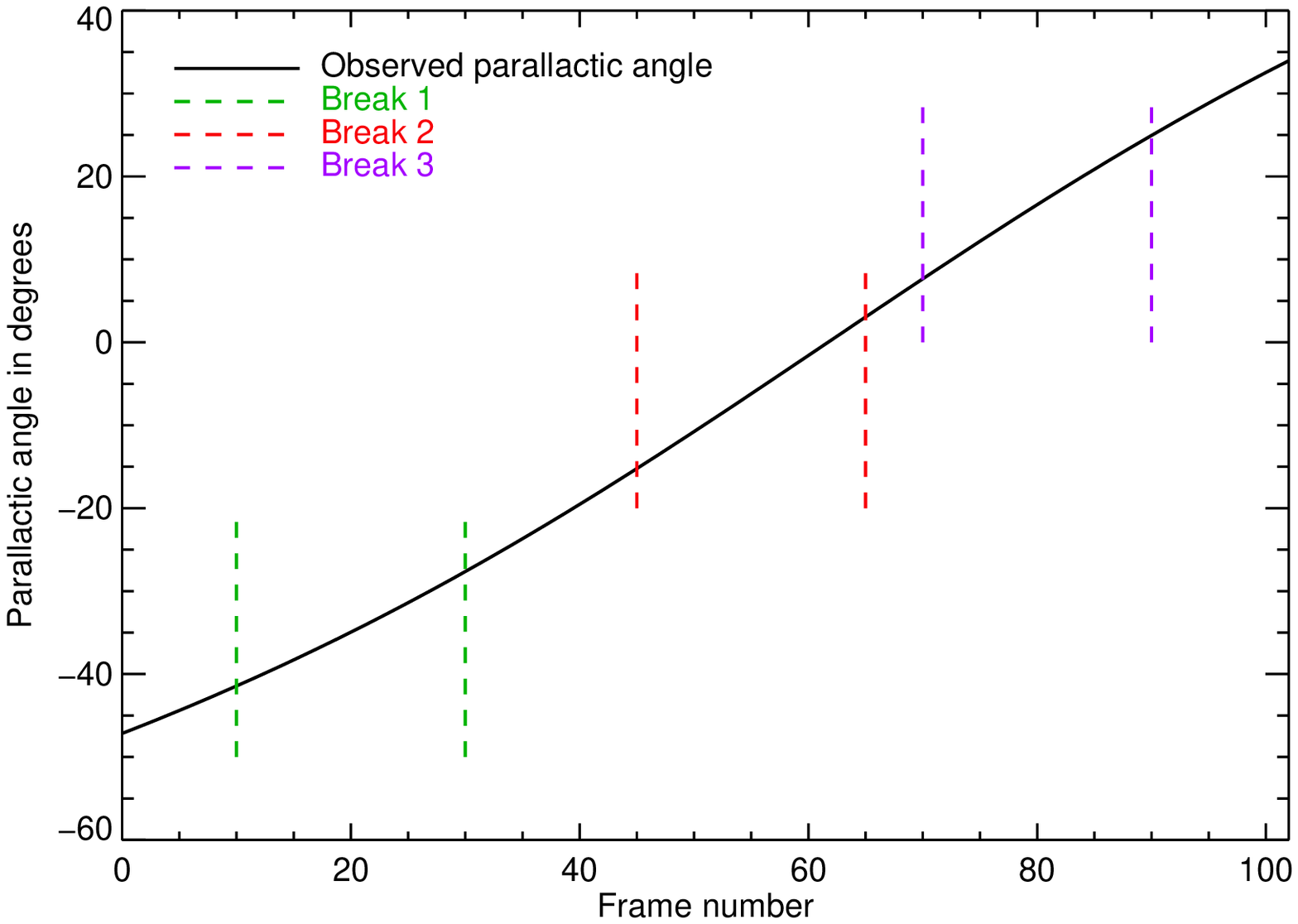}
   \includegraphics[width=8cm]{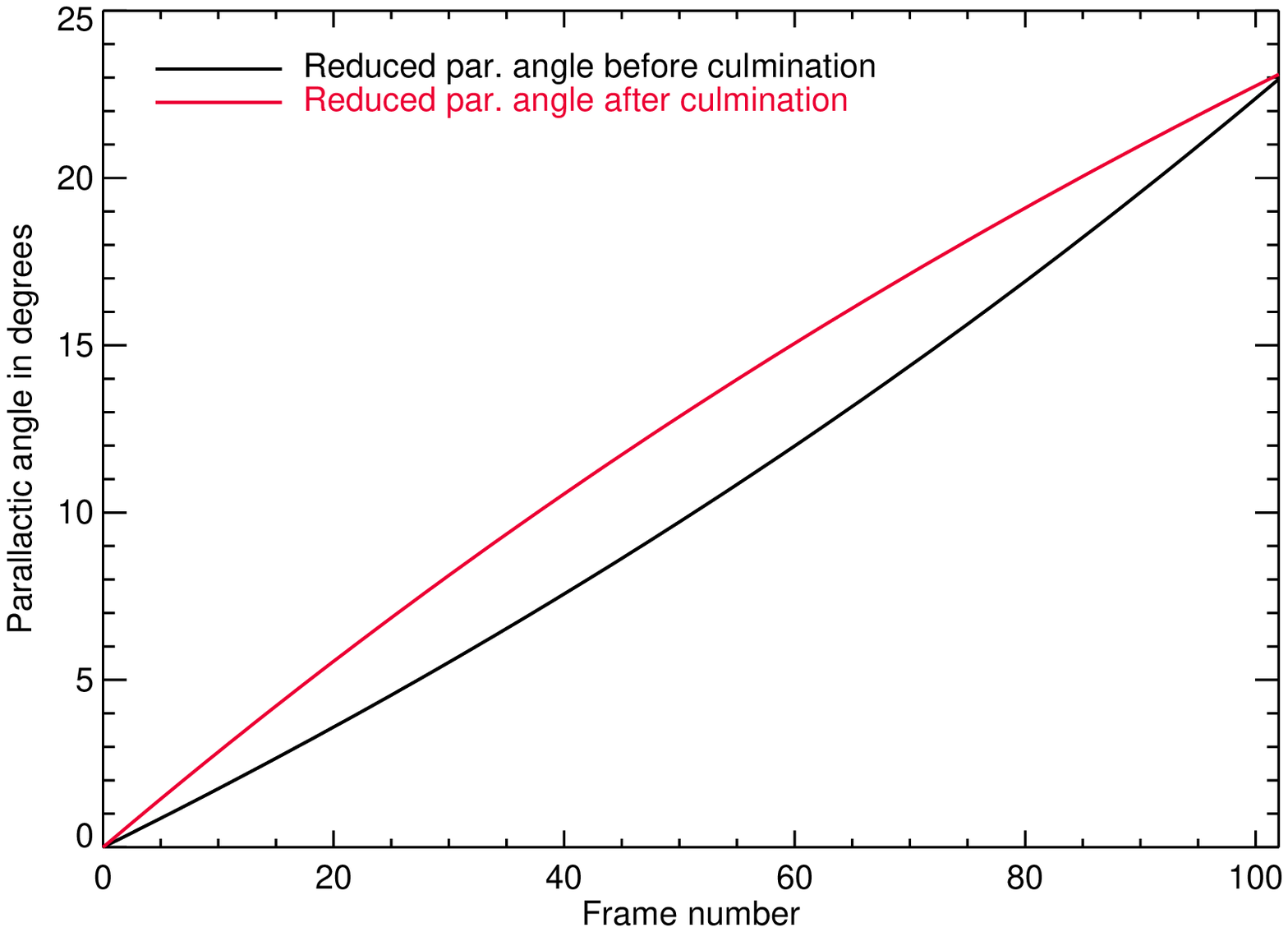}   
   \caption{Top: parallactic angle evolution of HR4796 as observed on April, 7 2010. The dashed lines show the artificial discontinuities introduced for test purposes. Bottom: parallactic angle evolution for a total field of view rotation $\Delta \theta=23^\circ$ before and after the star culmination in the sky.}
         \label{FigParallacticDefinitions}
   \end{figure}

To highlight those effects, we reduced simulated images of the disk of HR4796 as if they were observed in the same conditions as in the real observations on April, 7 2010, detailed in \citet{Lagrange2012b}. The same field of view rotation ($77.5^\circ$) and the same number of images in the datacube (103) after sorting poor-quality images were used. The evolution of the parallactic angle for the 103 images of the data cube is plotted in Figure \ref{FigParallacticDefinitions} at the top.

We used five additional parallactic angle evolution patterns: 
\begin{itemize}
\item the same parallactic angles as previously but with a discontinuity corresponding to eliminating 21 images from the data cubes at three different positions within the cube (see Figure \ref{FigParallacticDefinitions} at the top.) This discontinuity can mimic data sorting due to poor atmospheric conditions or quality of the adaptive optic correction
\item a reduced field of view rotation of $23^\circ$ observed after culmination of HR4796 in the sky, as in \citet{Thalmann2011} (see Figure \ref{FigParallacticDefinitions} at the bottom.)
\item a reduced field of view rotation of $23^\circ$ observed before culmination
\end{itemize}
The initial disk is convolved with a synthetic PSF in $L'$. The reductions were performed with cADI, rADI and LOCI (with a separation criterion $N_\delta$ of 1.5 FWHM and 3 FWHM).

\subsubsection{General effects of cADI}
\label{Sect:cADIeffects}

   \begin{figure}
   \centering
   \includegraphics[width=9cm]{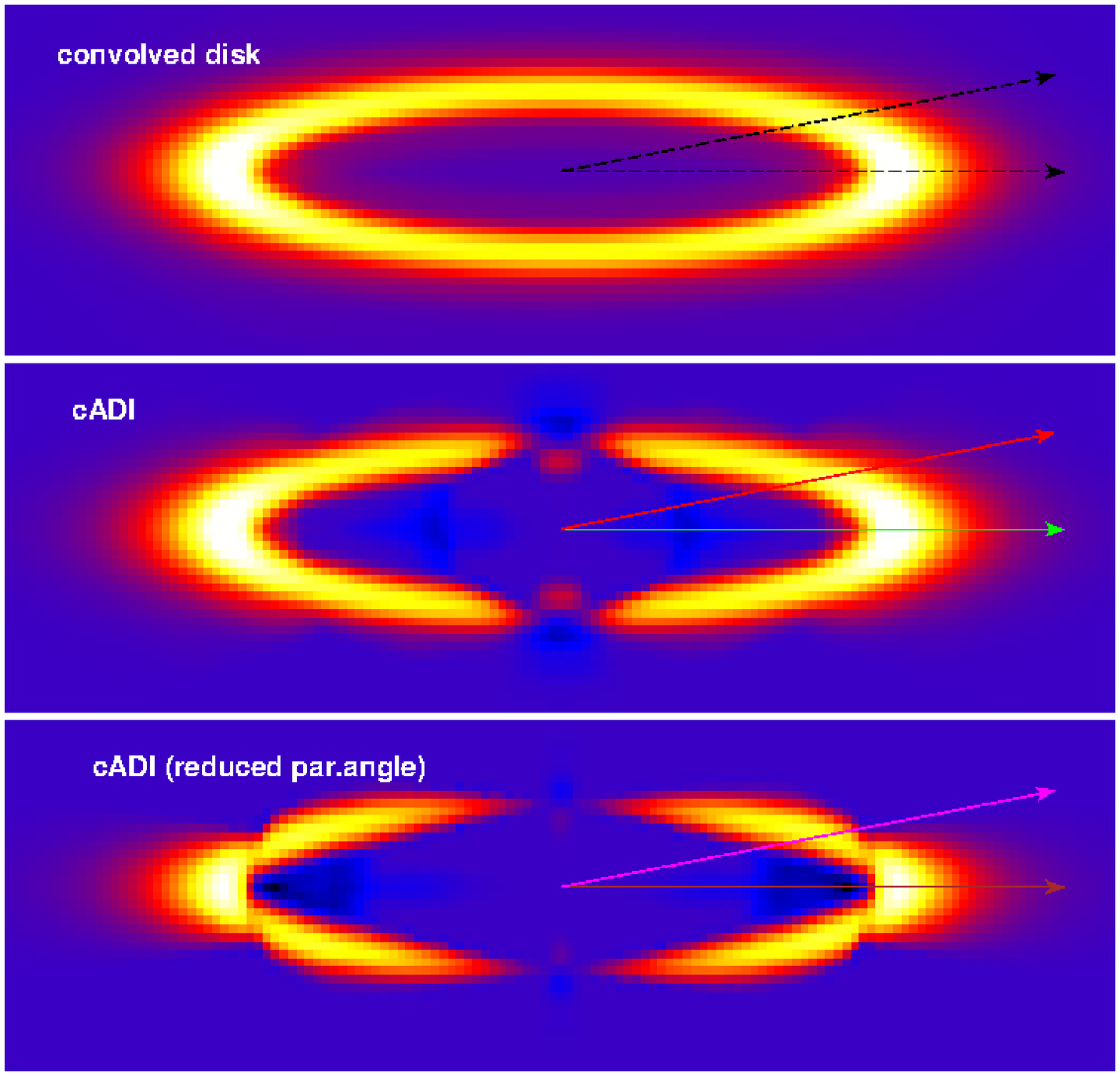}
   \includegraphics[width=9cm]{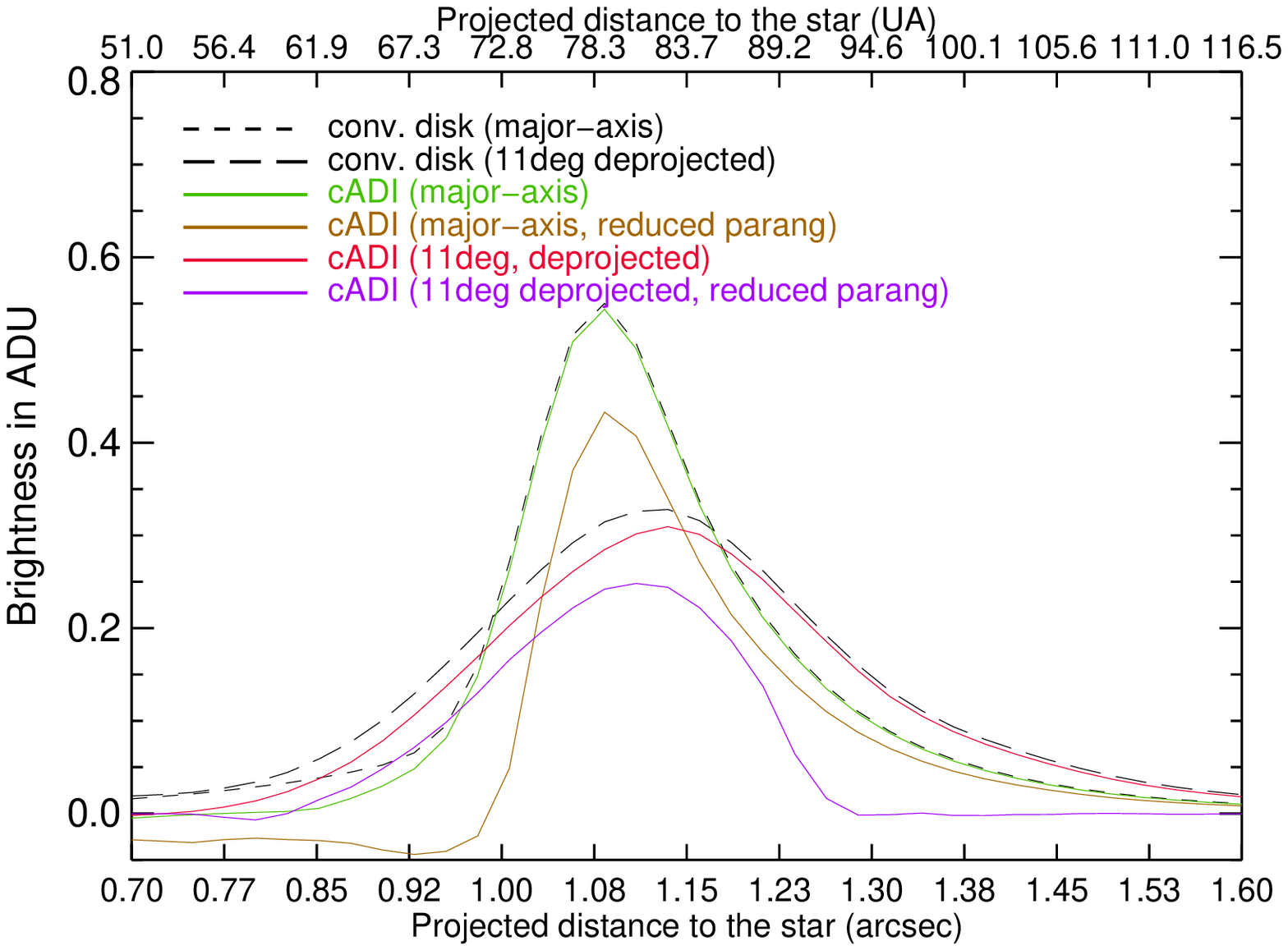}
   \caption{Top: initial convolved disk and cADI-reduced disks for the real parallactic angle evolution ($\Delta \theta=77.5^\circ$) and for the reduced amplitude ($\Delta \theta=23^\circ$). The color scales are identical and linear. Bottom: radial deprojected brightness profiles for the initial convolved disk (dashed lines) and the cADI images (plain lines) in two directions (along the semi-major axis and $11^\circ$ away from it). The color of the lines is the same as the color of the arrows in the top images.}
         \label{FigImagesArtefactsCADI}
   \end{figure}

   \begin{figure}
   \centering
   \includegraphics[width=6cm]{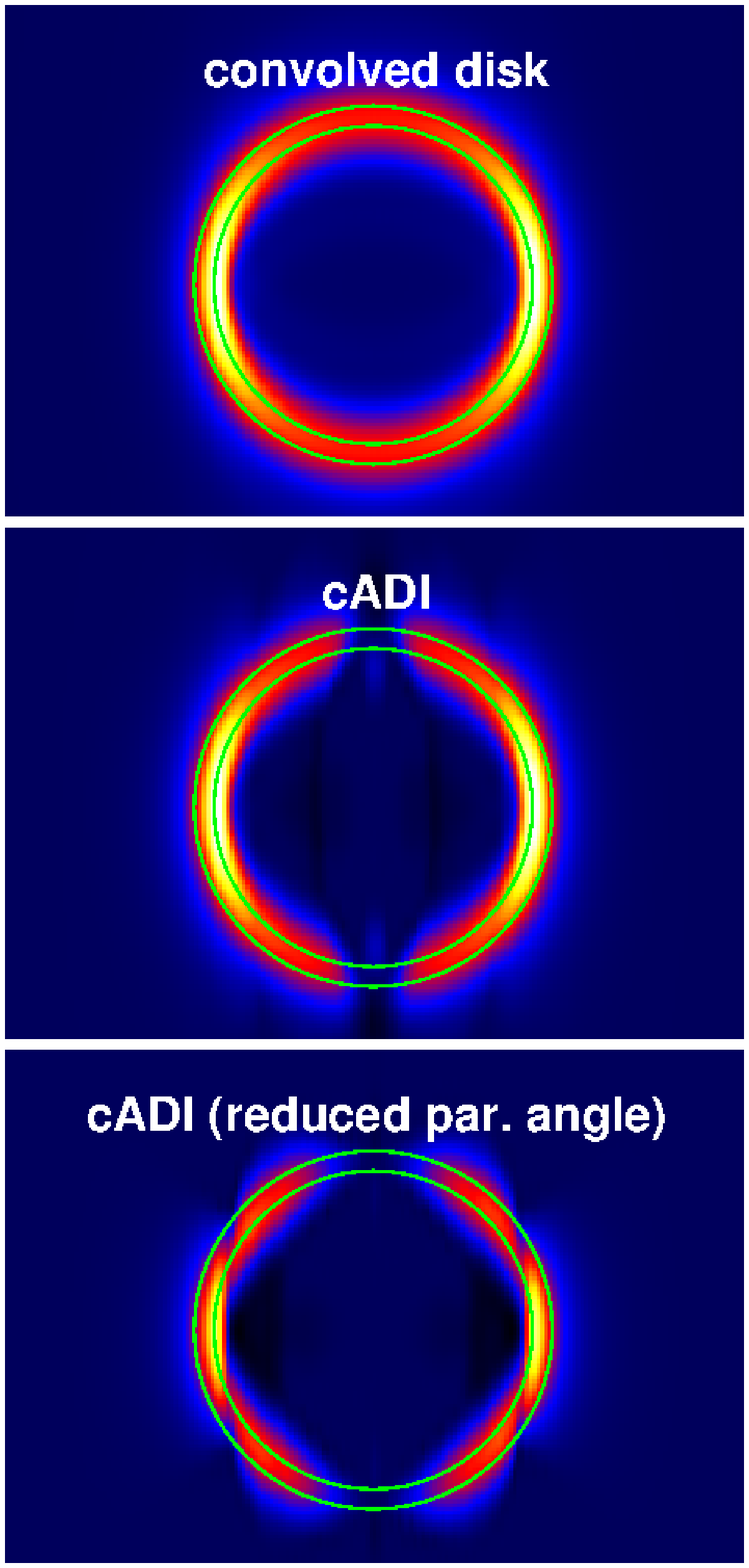}
   \includegraphics[width=9cm]{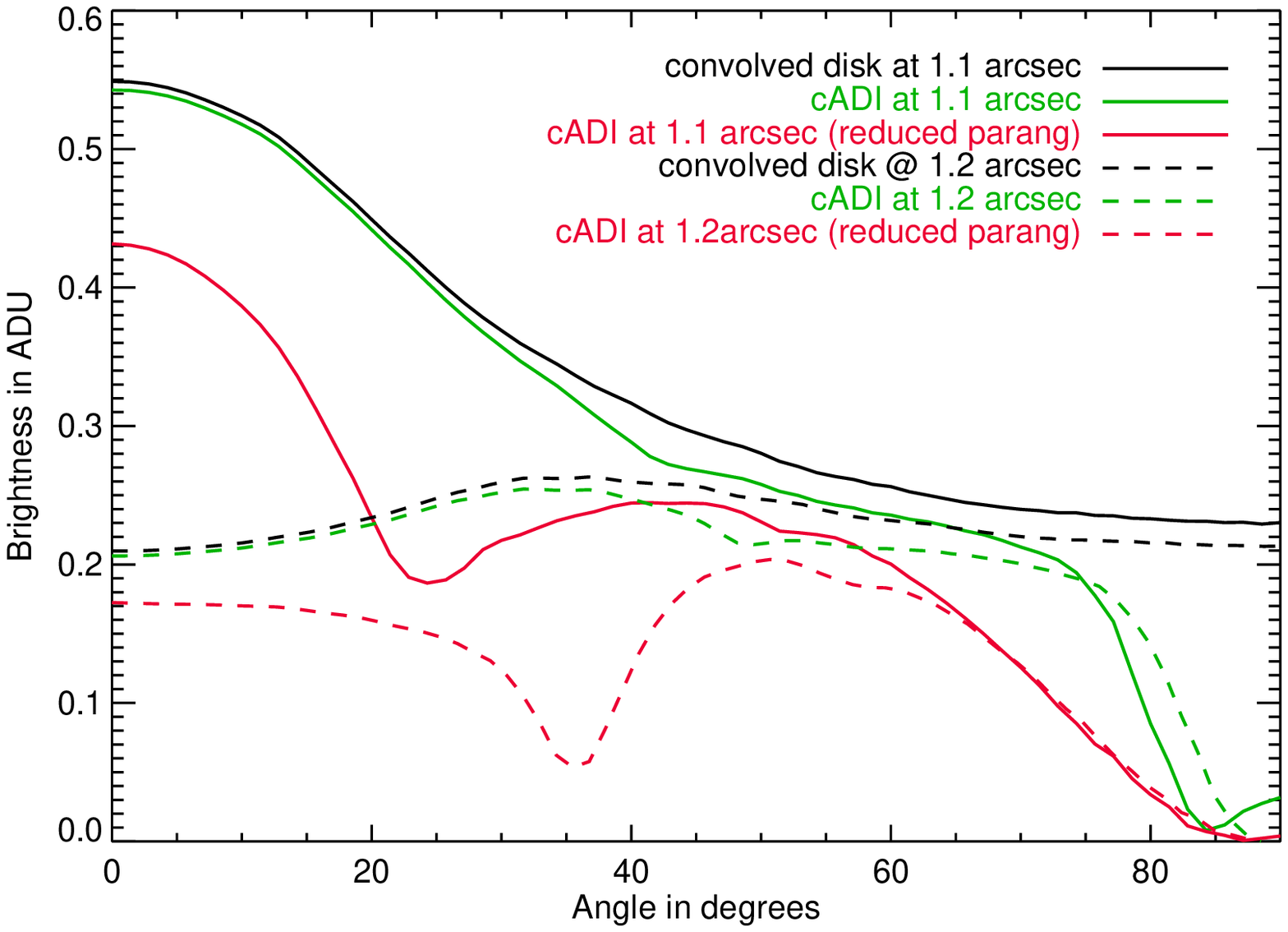}
   \caption{Top: deprojected disk. The two circles have a radius of $1.1\arcsec$ and $1.2\arcsec$. Bottom: azimuthal brightness profiles for the initial convolved disk, and the 2 cADI images along those two circles of radius $1.1\arcsec$ and $1.2\arcsec$. An azimuth angle of $0^\circ$ corresponds to the right in the top images.}
         \label{FigImagesArtefactsCADIdeprojected}
   \end{figure}

The first striking effect is the extinction along the minor axis of the disk seen in Figure \ref{FigImagesArtefactsCADI}. At this distance, the amplitude of field rotation is low and self-subtraction is significant, resulting in a progressive decrease in the width of the annulus as we approach the minor axis. 

With a $77.5^\circ$ field rotation, no other artificial  feature appears, except for a slightly negative region inside the annulus and two point-like sources along the minor axis with a flux $91\%$ lower than the ansae.
Nothing distinguishes this image from those with discontinuous parallactic angle evolutions: the introduced breaks have undetectable effects in cADI \footnote{down to a detection level of $0.1\%$ with respect to the ansae intensity}.   

For a reduced field rotation, two prominent artifacts appear for the northeast and southwest ansae of the annulus (left and right side in Figure \ref{FigImagesArtefactsCADI}, respectively). There are two extinctions on each side of the ansae, resulting in two blobs on the parts of the annulus that are most distant from the center. Applying the frequency-tool described in the previous section shows indeed that the azimuthal spatial frequencies at the location of the extinctions are of the same order of magnitude as the cutoff frequency due to ADI-filtering, which causes the enhanced flux losses in that region. 

As  mentioned in \citet{Lagrange2012b}, there is no need to involve radiation pressure-induced dust blow-out to explain the appearance of these bright structures named ”streamers” by \citet{Thalmann2011} in their high S/N data along the semi-major axis. The radial profiles along the semi-major axis at the bottom of Figure \ref{FigImagesArtefactsCADI} is unambiguous: the initial outer edge profile (small black dashed line) is well reproduced after cADI reduction (green line) even with a reduced parallactic angle amplitude (brown line) along the major-axis \footnote{A reduced parallactic angle amplitude attenuates the outer edge intensity by a uniform factor of $18\%$ but does not alter the shape and logarithmic slope of the profile along the semi-major axis. Only the inner edge is altered because of the presence of a negative area with a minimum intensity at $0.95\arcsec$. This conclusion does not hold any more in a direction $11^\circ$ away from the semi-major axis: in that case even the outer edge radial profile is altered, as shown by the violet curve in Figure \ref{FigImagesArtefactsCADI}.}.  

A deprojected view of the disk is helpful to understand where the cADI azimuthal profile deviates from the original disk profile (Figure \ref{FigImagesArtefactsCADIdeprojected}). The extinctions described quantitatively below are shown here by the local minimum in the azimuthal profiles (plain and dotted red lines). They are not present with a $77.5^\circ$ field rotation (green lines).

\subsubsection{Effects specific to rADI}

   \begin{figure}
   \centering
   \includegraphics[width=9cm]{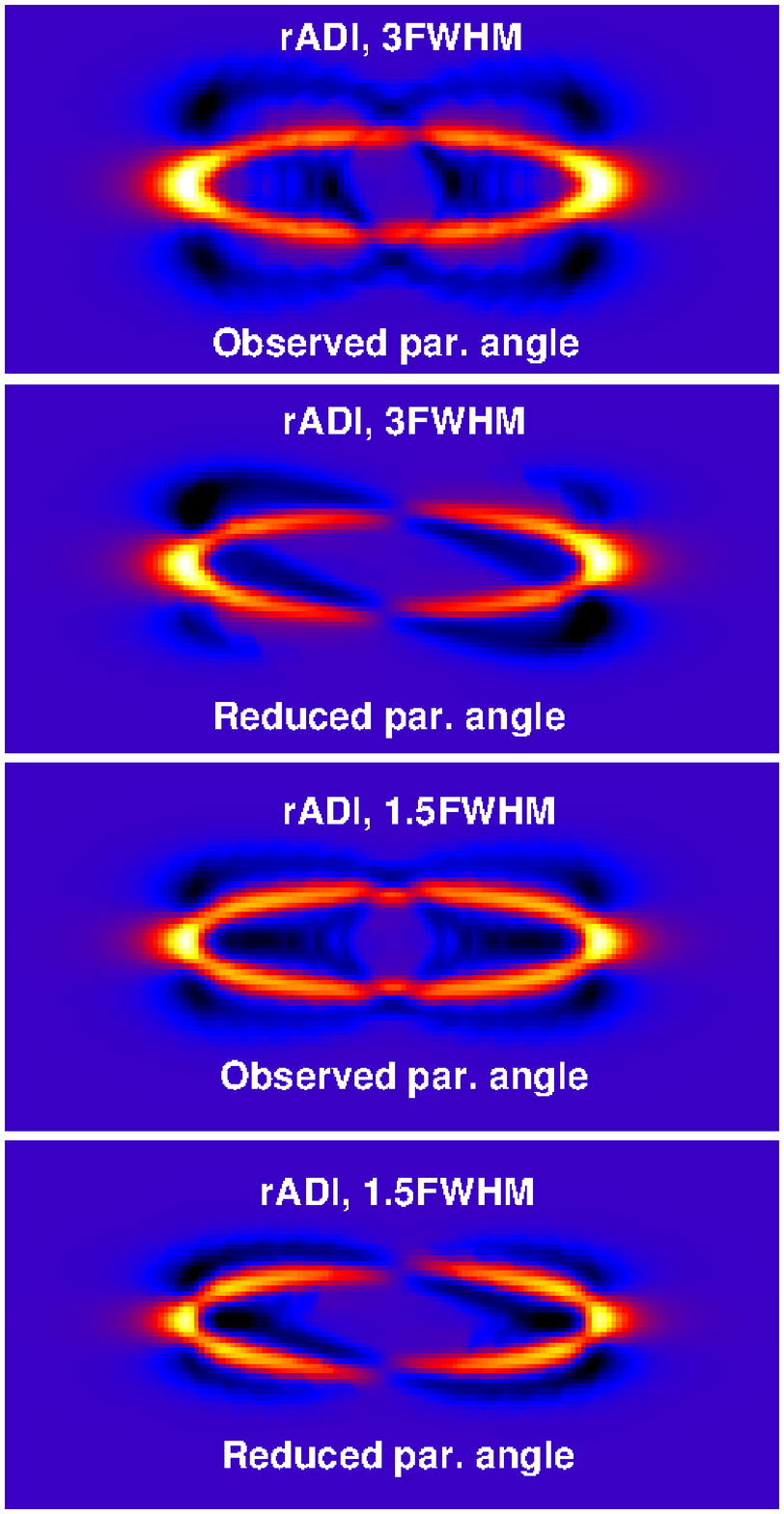}
   \caption{Residual disks after rADI processing for the observed parallactic angle evolution ($\Delta \theta=77.5^\circ$) and the one with reduced amplitude ($\Delta \theta=23^\circ$). The separation criterion $N_{\delta}$ is 3 FWHM for the 2 images at the top and 1.5 for the 2 images at the bottom. The color scales are identical and linear.}
         \label{FigImagesArtefactsRADI}
   \end{figure}

In rADI (Figure \ref{FigImagesArtefactsRADI}), the annulus appears thinner than in cADI. The blobs are still present at the tip of the ansae. The negative regions are emphasized. The differences between the two sets of parallactic angle evolution, observed and reduced amplitude, are not as high as in cADI, because the criterion that rules the reference frame selection is now $N_{\delta}$ rather than $\Delta \theta$. We see two negative regions on each side of the ring, looking like two negative rings inclined with respect to the positive ring. These two negative regions cross the ring close to the minor and major axis and create truncatures, as in cADI. The intensity of these negative regions is significant: it reaches $50\%$ of the ansae intensity for the reduced parallactic angle amplitude and $30\%$ for the real ($77.5^\circ$) amplitude. The angle between the negative and positive rings is directly driven by the separation criterion $N_\delta$ and the number of images $N$ used to build the reference PSF. The smaller these parameters, the closer the negative regions will appear with respect to the positive ring and the closer to the semi-major axis are the truncatures.

   \begin{figure}
   \centering
   \includegraphics[width=7cm]{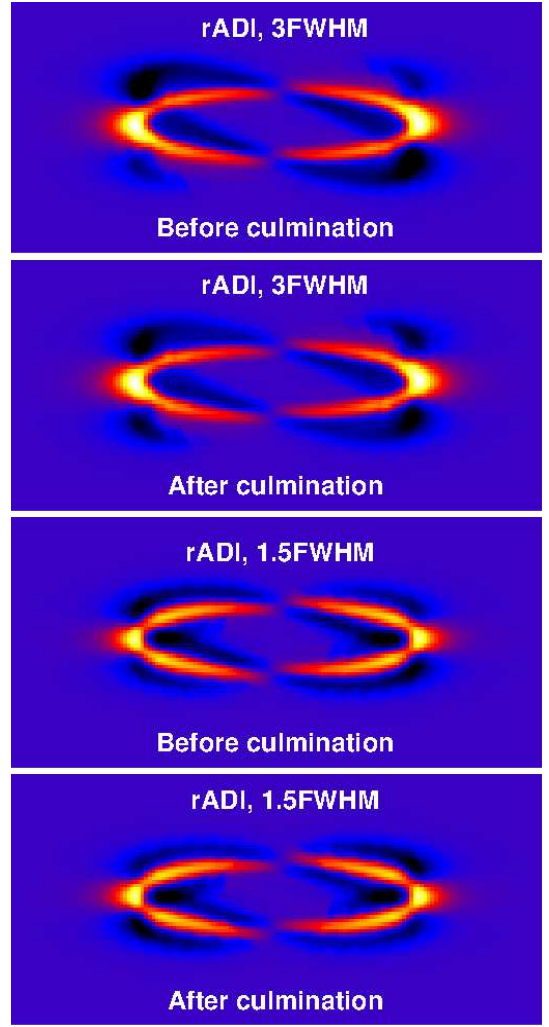}
   \caption{Disks after rADI processing for the so-called "reduced" parallactic angle evolutions ($\Delta \theta=23^\circ$). They reveal asymmetries depending on observing conditions before/after meridian. The separation criterion $N_{\delta}$ is 3FWHM for the two left images and 1.5 for the two images on the right. The color scales are identical and linear.}
         \label{FigImagesArtefactsRADICulminationEffect}
   \end{figure}

Unlike cADI, rADI images are variable whether we observe the star before or after culmination when it passes the meridian. This creates a left/right and top/bottom asymmetry as shown in Figure \ref{FigImagesArtefactsRADICulminationEffect}. This is because the reference frame is built using a reduced set of $N$ images in rADI, here 20. The negative regions appear to be stronger on the right side when the star is observed before culmination, e.g. when the parallactic angle evolution rate is increasing. This effect is stronger for a small separation criterion. 

   \begin{figure}
   \centering
   \includegraphics[width=7cm]{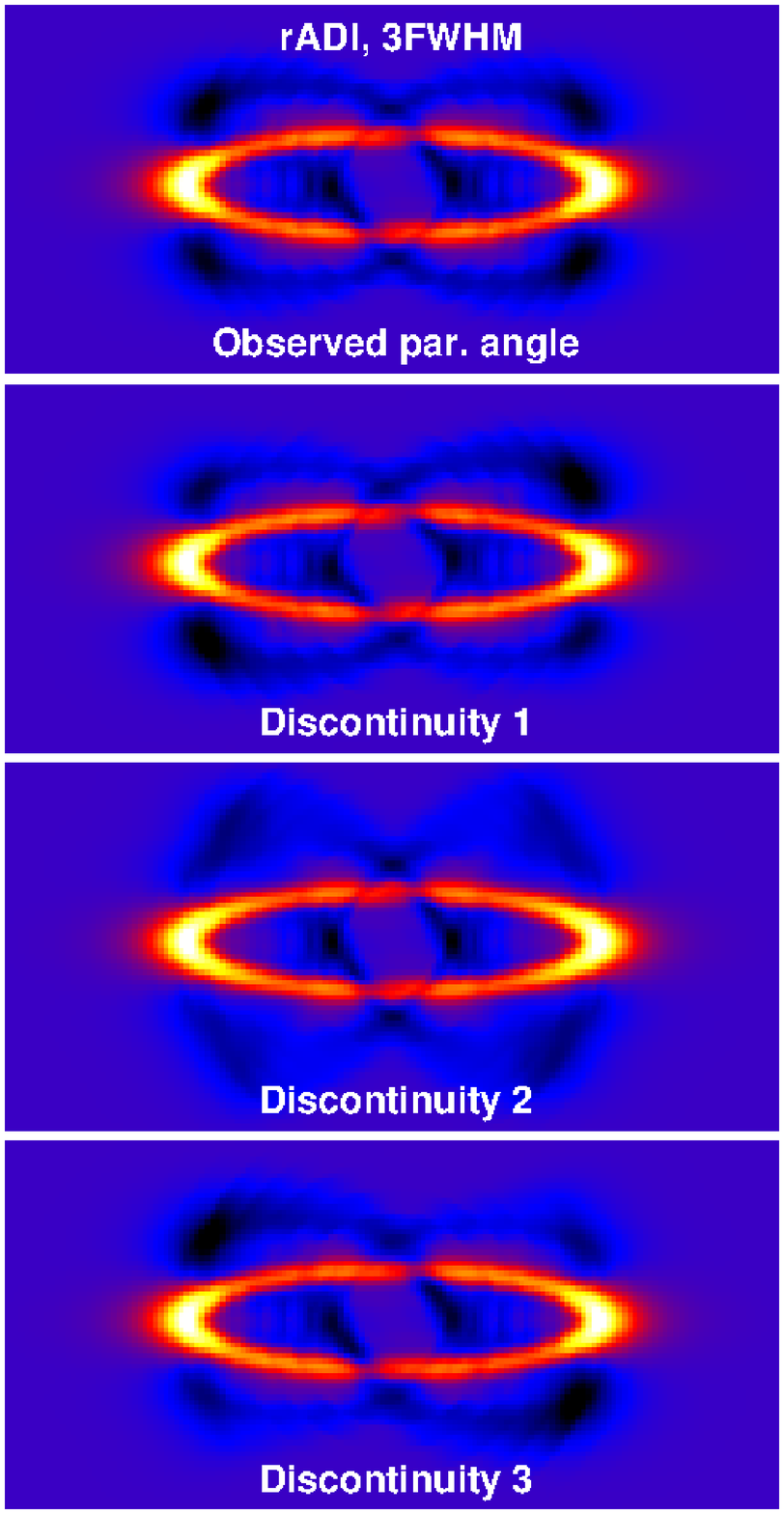}
   \caption{Disks after rADI processing for the observed parallactic angle evolutions ($\Delta \theta=77.5^\circ$) and for the three different discontinuities introduced. The separation criterion $N_{\delta}$ is 3FWHM for the two left images and 1.5 for the two images on the right. The color scales are identical and linear.}
         \label{FigImagesArtefactsRADIBreakEffect}
   \end{figure}

Discontinuity in the parallactic evolution pattern is another effect that create asymmetries detectable in rADI, as shown in figure \ref{FigImagesArtefactsRADIBreakEffect}. It easily induces brightness asymmetries in the negative regions up to $30\%$. The two negative regions are indeed symmetrical about the ring for the real observed parallactic angle evolution, but this is no longer the case when a discontinuity is introduced at the beginning (discontinuity 1) or at the end (discontinuity 3) of the sequence. When introduced in the middle of the sequence (discontinuity 2), the residual disk remains symmetrical. 

\subsubsection{Effects specific to LOCI}

In cADI or rADI, the way reference frames are built depends only on geometrical parameters. However, in LOCI, the linear combination also depends on the underlying speckle noise distribution. As a result, a relevant analysis of disk images reduced in LOCI cannot be performed without noise in the data. 

For that purpose, we used part of our $L'$ data of HR\,4796 from April, 6 2010 \citep{Lagrange2012b}, in which we inserted a fake disk with a S/N of about $20\sigma$ to test a few realizations of the LOCI optimization algorithm. The data cube consists of 67 images with $52^\circ$ field rotation.  Our goal here is not to make a comprehensive study testing all possible LOCI parameter sets and noise distributions, but to provide some hints on artificial effects occurring with LOCI.

\begin{figure}
   \centering
   \includegraphics[width=9cm]{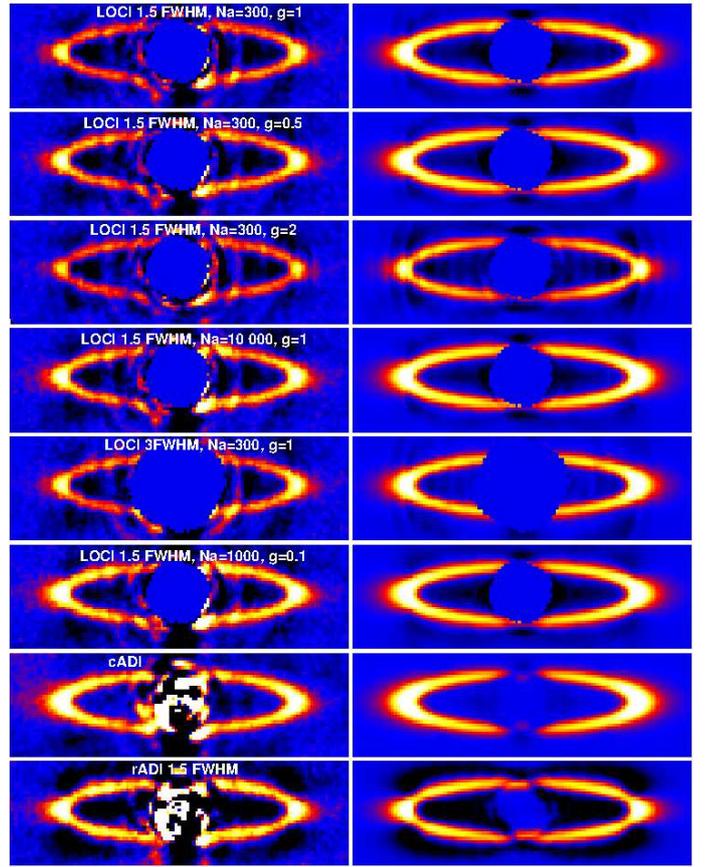}
   \caption{Left: residual disks after LOCI processing of the real data cube of HR4796 taken on April, 7 2010 with a fake disk inserted at $90^\circ$ from the real one. Different LOCI parameters are used and the coefficients are stored. Right: residual disk after applying those exact same LOCI coefficients to a data cube containing only the fake disk (without noise).The color scales are identical and linear for all images. The cADI and rADI reductions are shown at the bottom for comparison.}
         \label{FigImagesArtefactsLOCI}
   \end{figure}

For the baseline LOCI reduction, we used the following parameters: $N_\delta=1.5\times FWHM$, $N_A=300$, $g=1$ and $dr=1.4\times FWHM$. To distinguish between effects due to the inherent noise of the data and effects due to the LOCI algorithm, we stored for each reduction the LOCI coefficients $c^k$ used to build the reference frames and applied them to a data cube containing the fake disk without noise. We show in Figure \ref{FigImagesArtefactsLOCI} the real LOCI result on the left and the reconstructed disk image if there was no noise in the data on the right, for each set of LOCI parameters. 

As in rADI, all sets of parameters show bright blobs in the ansae when the separation criterion $N_\delta$ is small. The ring appears thinner elsewhere. The S/N is poorer close to the minor axis, therefore the noise can create distortion or intensity fluctuations in the ring that are not visible in the reconstructed image. 

No significant brightness asymmetry between the left and right ansae is detectable in the reduced LOCI images. Negative areas appear around and inside the ring, up to a level of $20\%$ of the flux in the ansae. The negative regions are more enhanced on the right side. 

The radial-to-azimuthal aspect ratio g appears to have an effect on the intensity and width of the ring. Optimization areas more azimuthally extended (i.e. with lower g) result in less self-subtraction. This is also the case for the set of conservative LOCI parameters \citep{Thalmann2011}, when $N_A=10000$, since this configuration means using the entire radial extension available in the image for each optimization area. The disk is $14\%$ brighter in that case, but negative areas are also emphasized. 

The highest flux in LOCI restitution occurs when a low aspect ratio g (here, 0.1 which is the same as using a full annulus) is combined with many PSF cores ($N_A$=1000). In this case, the disk intensity is almost similar to that obtained in cADI, except for the negative areas in the center of the ring.

It also turns out from these LOCI-reduced images that it is difficult to predict with the LOCI algorithm. 

\subsubsection{ADI applicability with respect to the disk inclination}

\begin{figure}
   \centering
   \includegraphics[width=9cm]{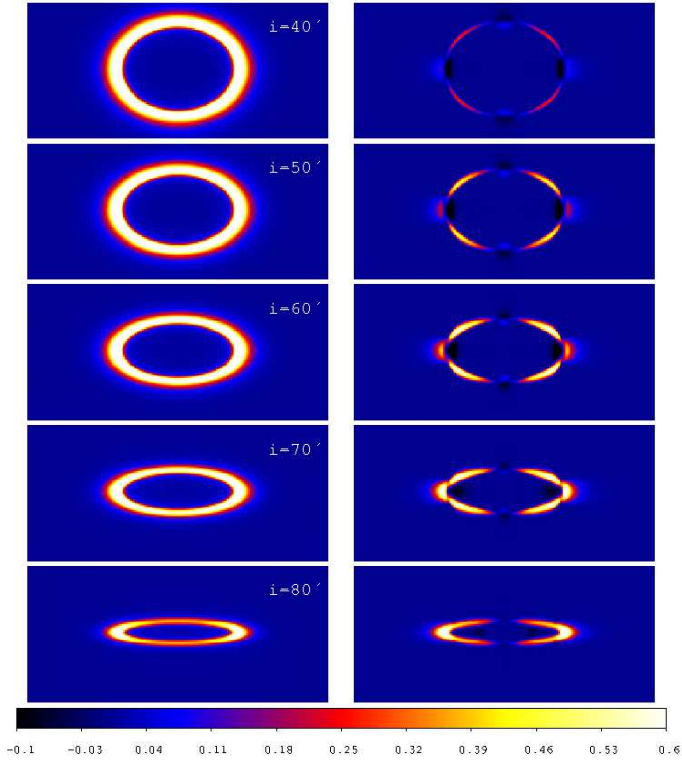}
   \caption{HR4796-like annuli for different inclination (left) and the residual disk image after cADI for a total field rotation of $50^\circ$. The color scale is identical and linear for all images.}
   \label{FigHR4796InfluenceInclination}
   \end{figure}

So far, we have studied both the flux loss and the ADI artifacts with disks that have a fixed inclination.
To answer the question to what extent in terms of disk inclination ADI can be applied, we performed additional cADI reductions of images of an inclined ring for several inclinations, as shown in Figure \ref{FigHR4796InfluenceInclination}. 
Obviously, no clear answer can be provided since many additional aspects must be considered, such as the disk parameters to be retrieved and the amount of self-subtraction acceptable given the data S/N.
However, we show here that for low-inclination, e.g. below $60^\circ$, the major and minor axis are critical regions affected by ADI because many low spatial frequencies are lost in those regions. Even for higher inclinations, the minor axis is always too much affected by self-subtraction to allow any information retrieval. Concerning the major axis, blobs at the tip of the ring ansae (described in more detail in Section \ref{Sect:cADIeffects}) appear below $70^\circ$ therefore great care must be taken not to misinterpret those features with physical processes such as dust grains blown by radiation pressure. These results suggest that disks inclined below $50^\circ$ are inappropriate for ADI. 

\subsection{ADI on spiral arms}

Recent disks discoveries included spiral arms in circumstellar disks. For instance, \citet{Fukagawa2006} performed reference star PSF subtraction to reveal spiral arms around HD142527, and \citet{Rameau2012} confirmed these features using both ADI and reference star subtraction. Other high-contrast imaging technique such as polarimetry were used for SAO206462 by \citet{Muto2012}.

Because spiral arms have a very specific geometry with azimuthal spatial frequencies varying with the separation from the star, we carried out a specific investigation of the ADI impact on these features. 

   \begin{figure}
   \centering
   \includegraphics[width=9cm]{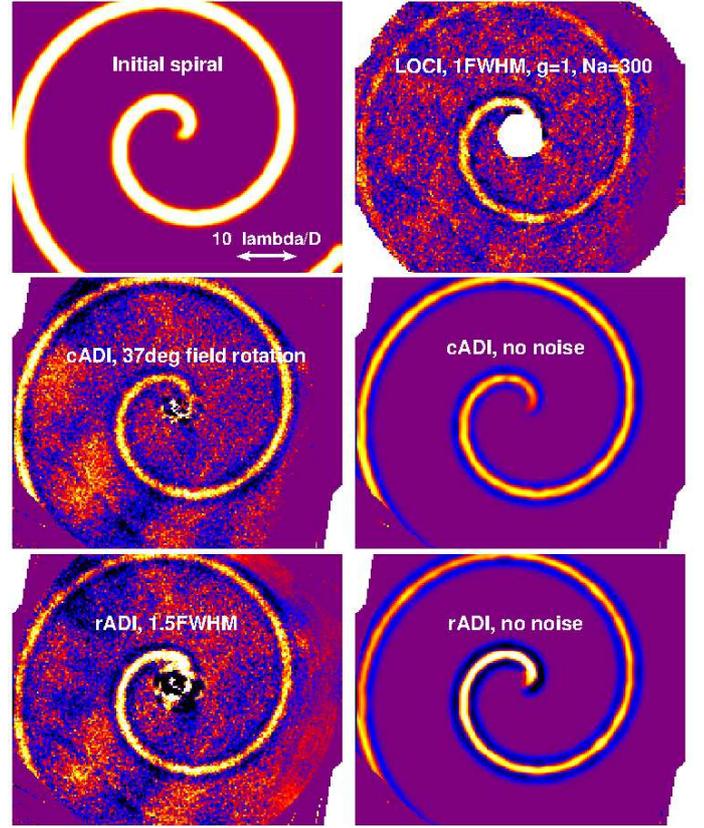}
   \caption{ADI reduction of a spiral arm seen pole-on, with and without noise from L' data. The color scales are identical and linear for all images.}
         \label{FigImagesArtefactsSpiral}
   \end{figure}
   
An Archimedian spiral, seen pole-on, parametrized in polar coordinates by $r=10\times \theta$ with an initial width of 10 pixels was processed through cADI, rADI and LOCI with a total field rotation $\Delta\theta = 37^\circ$. Results are shown in Figure \ref{FigImagesArtefactsSpiral}. To allow a relevant comparison of cADI and rADI images with LOCI images, we added noise to the data. For that purpose we used our L' data (FWHM=3.6px) of the M star HR\,4796B taken on April, 6 2010 \citep{Lagrange2012b} in which we inserted the spiral arm with an S/N of 14.

The first conclusion is that arms are significantly thinner. The width reduction reaches $30\%$ in cADI. 

As previously, we observe negative shadows on each side of the arm. These shadows can be deep and not necessarily symmetrical about the arm. This is mostly visible in rADI without noise. When the curvature is smaller than the curvature of a circle at the same location (e.g. for the innermost regions), the negative parts on the outer edge of the spiral are enhanced, and vice versa. 

Great care should also be taken when interpreting the intensity distribution of a spiral arm. Although the initial spiral has a uniform flux, the LOCI and rADI reductions show a higher flux at short separation. The flux loss due to self-subtraction is indeed a trade-off between how large the amplitude of field rotation is and how small the azimuthal component of the spatial frequencies is.  

The case of a spiral arm inclined towards the line of sight was also investigated. It appears to be very similar to the case of an inclined ring. Some regions behave like the semi-major axis of an inclined ring, creating bright blobs when the amplitude of field rotation is low and some regions behave like the semi-minor axis of an inclined disk where most the disk flux is self-subtracted.

\section{Quantitative estimation of the biases}
\label{SectionQuantitative}

We now provide a quantitative estimation of the change in some important disk observable properties from an astrophysical perspective: the disk total flux,  the outer slope of the SBD and the ring full width at half maximum (FWHM) measured radially along the semi-major axis. For that purpose, we focus on ring-shaped disks. We analyze the effect of the following variable parameters: the inclination of the ring, the radial exponents $\alpha_{out}$ and $\alpha_{in}$ that parametrize the volumetric dust density for three different ADI algorithms as detailed in Table \ref{TabParameterSet}. 

\begin{table}[t!]
\caption{Set of disks and ADI parameters used for the quantitative analysis.}
\label{TabParameterSet}
\begin{center}
\begin{tabular}{c|c}
$\Delta \theta$ & $52^\circ$ \\
\hline
i & $60^\circ$;$65^\circ$;$70^\circ$;$75^\circ$;$80^\circ$;$85^\circ$ \\
\hline
$\alpha_{in}$ & 2;3;5;7;10\\
\hline
$\alpha_{out}$ & -2;-3;-5;-7;-10\\
\hline
\multirow{3}{*}{ADI algorithm} & cADI \\
& rADI $N_\delta=1.5$, $N=20$, $dr=1.4$ \\
& LOCI $N_\delta=1.5$, $g=1$, $N_A=300$, $dr=1.4$ \\
\end{tabular}
\end{center}
\end{table}

For each set of disk parameters and ADI algorithms, the reductions were performed with and without noise. For the case with noise, we again used part of our $L'$ data of HR\,4796A from April, 6 2010 \citep{Lagrange2012b}, into which we inserted a fake disk with a S/N of about $10\sigma$. The data cube consists of 67 images with $52^\circ$ field rotation.
 
The slope and FWHM measurements along the semi-major axis appear to be very sensitive to the noise, particularly for very narrow rings because we lack sufficient dynamic in the image. Therefore, to isolate the effect of ADI, we performed those two measurements on the images reduced without noise. For LOCI we computed the LOCI coefficients for data with noise and applied them to a noiseless model of the disk.

\subsection{ADI impact on the total flux of the disk}
\begin{figure*}
   \centering
   \includegraphics[width=15cm]{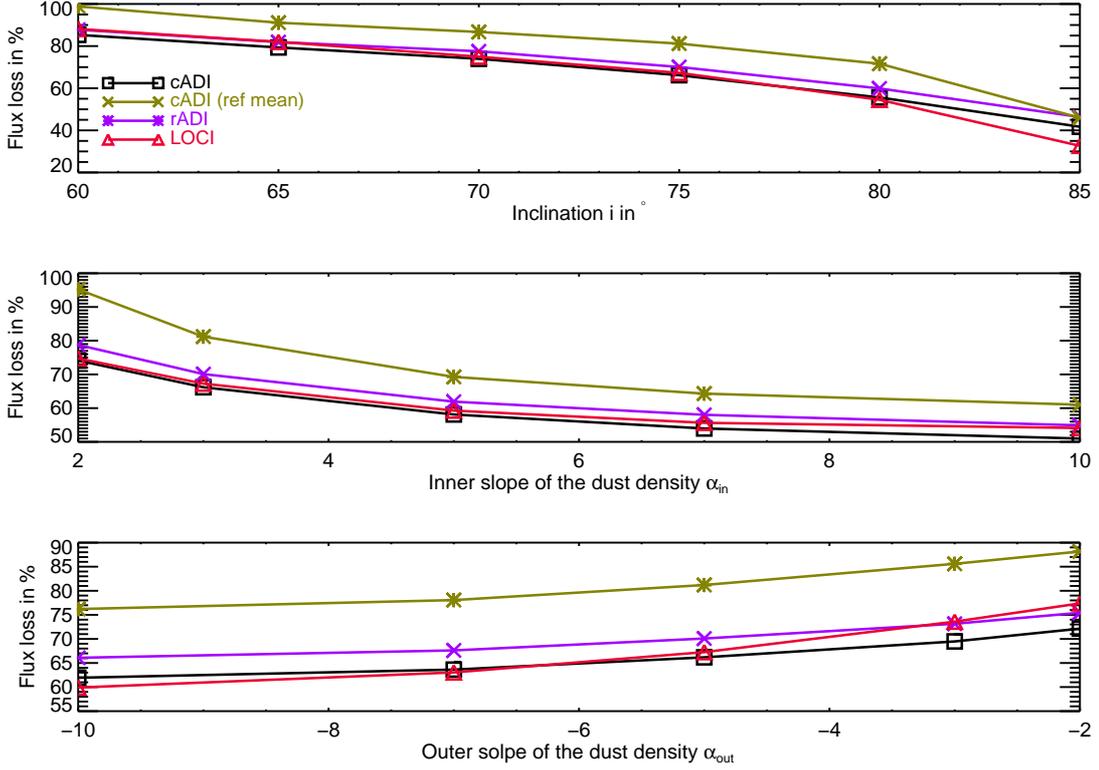}
   \caption{Flux loss on a disk as a function of the disk inclination i (top chart), the dust density inner slope $\alpha_{in} $(middle chart) and the dust density outer slope $\alpha_{out}$ (bottom chart). The flux loss is the ratio of the total flux of the reduced disk by the total flux of the initial disk after convolution by a synthetic PSF. For the top chart, $\alpha_{in}$ and $\alpha_{out}$ are kept constant to 3 and -5, for the middle chart i and $\alpha_{out}$ are kept constant to $75^\circ$ and -5, and for the bottom chart i and $\alpha_{in}$ are kept constant to $75^\circ$ and 3.}
   \label{FigChartTotalFlux}
   \end{figure*}

Following the general model detailed in Section \ref{SubsectionTheoreticalModel}, ADI behaves as a high-pass filter, therefore the smoother the disk brightness distribution, the greater the disk flux loss. We illustrate this in Figure \ref{FigChartTotalFlux} in a number of quantitative examples coming from the reduction tests cases detailed in Table \ref{TabParameterSet}:

\begin{enumerate}
\item The  amount of flux loss after ADI processing is very highly correlated to the inclination of the disk. The more pole-on the disk, the higher the flux loss, whatever the algorithm used. Figure \ref{FigChartTotalFlux} provides typical examples that can be used as references, with flux losses ranging from $100\%$ to a few $\%$ for disks inclined from $60^\circ$ to $85^\circ$. 
   
\item The inner and outer slopes of the dust density distribution also have a significant impact on the surface brightness of the processed disk, since the softer the slopes, the higher the flux loss. For the expected slopes of circumstellar disks, e.g. about -2 to -5 for $\alpha_{out}$ and 2 to 5 for $\alpha_{in}$, the flux loss varies between $50\%$ and $80\%$ , as visible in the middle and bottom chart of Figure \ref{FigChartTotalFlux}. 

\item In cADI, flux losses are about $10\%$ higher when the reference frame is built using the mean rather than the median. This conclusion contradicts the graphs of Figure \ref{FigCADIRegimes}, which were produced from noiseless data. The mean is indeed much more sensitive to noise, which leads to more disk flux captured in the reference frame in the presence of noise. 

\end{enumerate}

	\subsection{ADI impact on the outer slope of the surface brightness distribution}

Because the inner or outer slope of a debris disk is a significant parameter for understanding the physics of the disk, we studied the effect of ADI on the slope along the major axis of a ring. As already seen, the inner part of the ring is strongly affected by self-subtraction, therefore attempting to retrieve the dust density inner slope from ADI images is dangerous. For this reason, we focus here on the parameters that influence the outer slope of the SBD.

\begin{figure*}
   \centering
   \includegraphics[width=15cm]{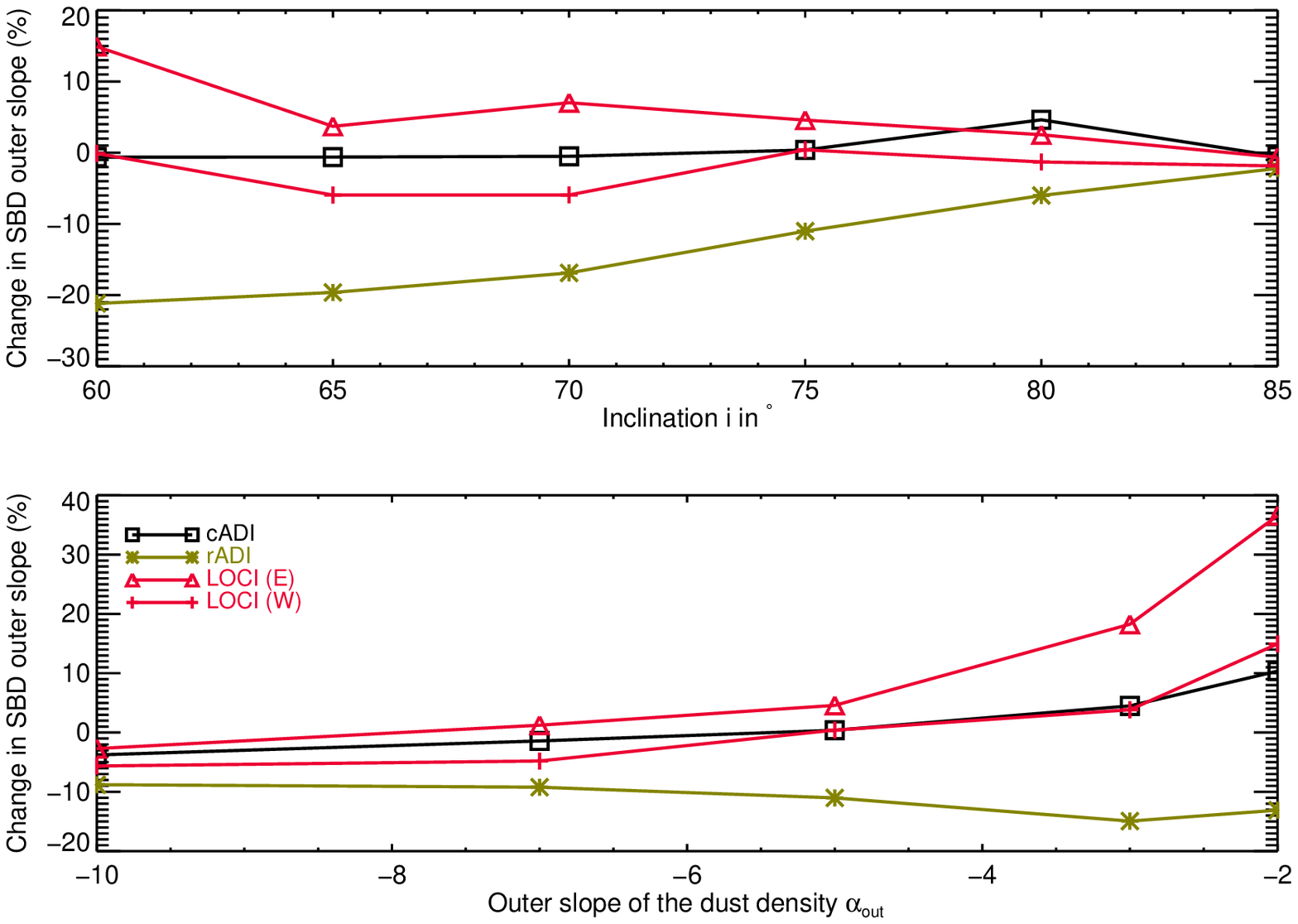}
   \caption{Bias on the measured outer slope of the disk SBD as a function of the disk inclination i (top chart) and the dust density outer slope $\alpha_{out}$ (bottom chart). For the top chart, $\alpha_{in}$ and $\alpha_{out}$ are kept constant to 3 and -5 and for the bottom chart, i and $\alpha_{in}$ are kept constant to $75^\circ$ and 3.}
   \label{FigChartImpactOuterSlope}
   \end{figure*}

The SBD outer slope is retrieved from a linear regression on a log-log plot of the SBD profile. Two elements have an impact on the slope of the final image: the convolution of the initial disk with the PSF that tends to make the slope of the profile softer, and ADI processing whose effect is non-linear. 
The relative change in the SBD outer slope due to ADI reduction is presented in Figure \ref{FigChartImpactOuterSlope}. A negative change means a steeper profile. As the reduced LOCI image is no longer symmetrical about the semi-minor axis, we provide here the slope measurements on each side of the semi-major axis. We show here the impact of the parameters i and $\alpha_{out}$ only, as the inner slope of the dust density $\alpha_{in}$ has no influence on the SBD outer slope.

\begin{enumerate}
\item in cADI (median-combined reference frame), the inclination and initial value of $\alpha_{out}$ has little impact on the slope. The slope is slightly softer for values of $\alpha_{out}$ close to -2. 
\item in LOCI, the effect is variable whether we consider the East or West side of the ring. The trend seems however to follow that of cADI.
\item in rADI, the SBD slope is clearly steeper, by -20 to -5\% for the explored range of inclinations and $\alpha_{out}$. 
\end{enumerate}
On the whole the effect is all the more pronounced as the disk is spatially extended with low inclination and soft SBD slope.

\subsection{ADI impact on the FWHM of the ring}

In almost all simulation cases, the measured FWHM is slightly decreased after ADI reduction.

   \begin{figure*}
   \centering
   \includegraphics[width=15cm]{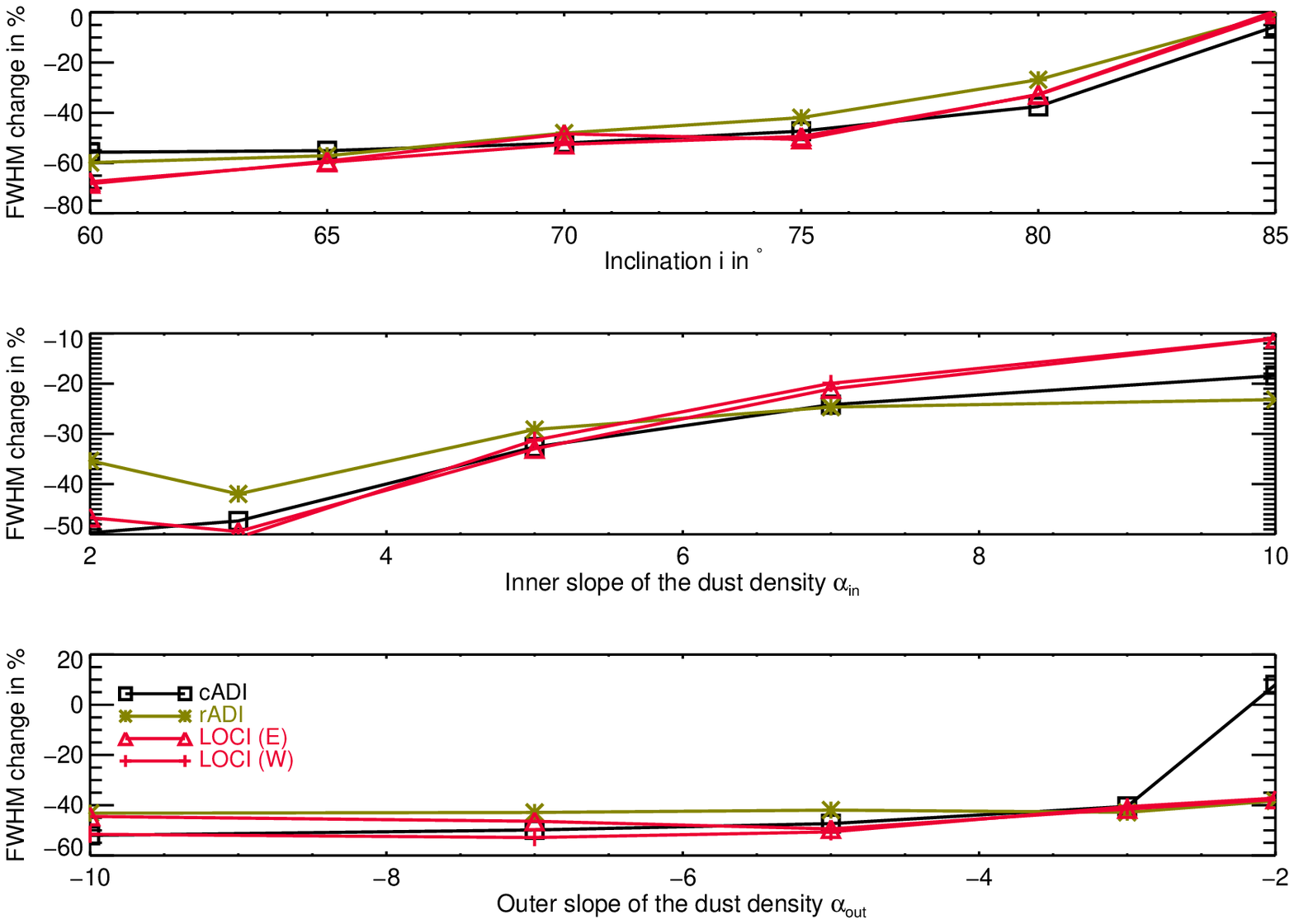}
   \caption{Bias on the FWHM of the ring (measured along the semi-major axis) as a function of the disk inclination i (top chart), the dust density inner slope $\alpha_{in} $(middle chart) and the dust density outer slope $\alpha_{out}$ (bottom chart). For the top chart, $\alpha_{in}$ and $\alpha_{out}$ are kept constant to 3 and -5, for the middle chart i and $\alpha_{out}$ are kept constant to $75^\circ$ and -5, and for the bottom chart i and $\alpha_{in}$ are kept constant to $75^\circ$ and 3.}
   \label{FigChartFWHM}
   \end{figure*}

\begin{enumerate}
\item First, the inclination of the disk can significantly bias the measured FWHM. The more face-on the disk is seen, the higher the impact on the FWHM compared to the initial convolved disk (see Figure \ref{FigChartFWHM} top chart).  
\item The measured FWHM of a ring is also highly dependent on the dust distribution inner slope of the initial disk. Indeed, for very soft radial density exponents $\alpha_{out}$, the initial FWHM of the ring is large. ADI then creates negative areas inside the ring, consequently the inner half-width is much decreased. This effect is visible in the middle chart of Figure \ref{FigChartFWHM}.
\item Last, the dust density outer slope has a fairly low impact on the disk FWHM. In rADI, the change in FWHM is constant over the explored range of $\alpha_{out}$ and the change decreases from -38\% to -22\% (without noise) when $\alpha_{out}$ approaches 0 (Figure \ref{FigChartFWHM}, bottom chart).
\end{enumerate}

\subsection{Sensitivity analysis relative to the LOCI parameters}

The study presented above compares the reduction of various initial disks but with a single set of rADI and LOCI parameters. To complete this study, we now show for a given disk morphology, namely $\alpha_{in}=5$,	 $\alpha_{out}=-2.5$ and $i=75^\circ$, a comparison of different LOCI parameters. 

\begin{figure*}
   \centering
   \includegraphics[width=15cm]{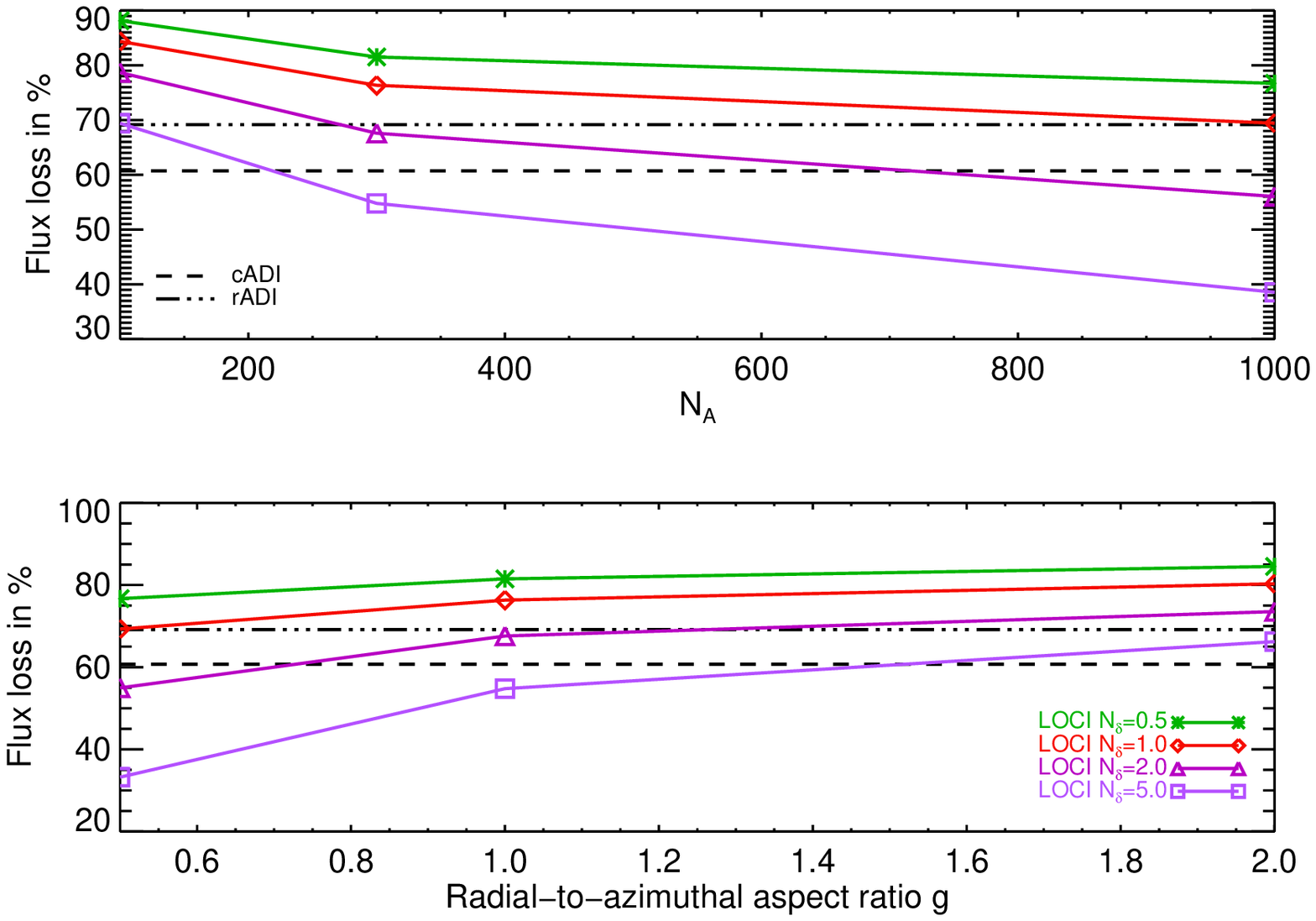}
   \caption{Flux loss on a disk as a function of the size of the LOCI optimization region $N_A$ (top chart) and its radial-to-azimuthal aspect ratio g (bottom chart). An initial disk parametrized with $i=75^\circ$, $\alpha_{in}=5$ and $\alpha_{out}=-2.5$ is used here, with g kept constant to 1. for the top chart and  $N_A$ kept constant to 300. for the bottom chart. Flux losses in cADI and rADI ($N_\delta=1.5$) are shown here for reference.}
         \label{FigChartLOCITotalFlux}
   \end{figure*}

We have already shown that the separation criteria $N_\delta$ is the main parameter that influences the amount of self-subtraction of a disk. The size of the LOCI optimization area, expressed in number of PSF cores $N_A$, is the second contributor. The larger this area, the lower the flux loss, as shown in the top chart of Figure \ref{FigChartLOCITotalFlux}.  For a large domain, the chances that the RMS optimization matches the disk pattern and not the speckle pattern is indeed lower.

Now, for a constant size of the optimization area (here $N_A$ is kept to 300), the aspect ratio is also influencing the amount of self-subtraction, as shown in the bottom chart of Figure \ref{FigChartLOCITotalFlux}. For the morphology studied, the larger the radial extension of this area, the less self-subtraction occurs. We interpret this feature by the fact that the relative area contaminated by the disk is lower for the optimization zones characterized by low values of g. This conclusion is obviously valid for the geometry of disks considered here and should be revisited for a disk different from an inclined ring.

\subsection{Position angle and offset of the disk}

The position angle (PA) and offset of a disk are very important parameters for describing the morphology of a disk. Within ADI, two different reasons can alter the PA and offset as measured on the reduced disk image:
\begin{itemize}
\item star centering
\item artifacts resulting from disk self-subtraction.  
\end{itemize}

Star centering is an important problem in ADI because image rotations or derotations are frequently performed, and one does not always accurately know the star center for coronographic or saturated images. The PA of a disk is less sensitive a parameter than the PA of a planet, however, as \citet{Lagrange2012} showed for the $\beta$ Pictoris disk and its planetary companion. In the case of HR\,4796A \citep{Lagrange2012b}, the disk PA uncertainty due to the imperfect knowledge of the NaCo PSF center reaches $0.2^\circ$, which is the same order of magnitude as the uncertainty associated to the PA measurement technique (the maximum regional merit method, as in \citet{Buenzli2010} and in \citet{Thalmann2011}). Concerning the the uncertainty on the disk center due to the uncertainty on the PSF center, we also showed in \citet{Lagrange2012} that it is the same order of magnitude as the uncertainty on the star center, 0.2 px in that case.

Then we investigated the influence of artifacts resulting from the disk self-subtraction on the PA and offset measurement. In \citet{Lagrange2012b}, we reduced simulated disk images of HR\,4796A in cADI, rADI and LOCI and compared the PA and offset of the reduced disk image to their inital value, as measured with the regional merit method. 
Two different initial disk models were used, called HR4796SD and HR4796blowoutSD, and $77.5^\circ$ field rotation was used in the simulation. In both cases, the error on the disk PA due to ADI was below $0.1^\circ$ or $0.1\%$. The measured bias on the offset of the disk was within the measurement uncertainty of the regional merit method.

We emphasize that some specific configurations might nevertheless bias the PA measurement. We see indeed in Figure \ref{FigImagesArtefactsRADI} that rADI reduction with $23^\circ$ field rotation tends to artificially create a discontinuity between the right and left sides of the ring that can bias the PA measurement if made separately for each side.

\subsection{Overestimation of planet detection limits}

It is well known that dust surrounding stars is a possible source of confusion with exoplanets \citep{Roberge2012} and can lead to false-positive detections. Even without detection, the presence of a disk can also lead to an overestimation of planet detection limits.

The detection limits are usually computed by first calculating the flux loss for a fake planet injected in the data and then applying this correction factor to an estimation of the $5 \sigma$ noise level in the reduced ADI image. Great care should be taken in this case because the region inside the annulus and around it has a negative flux, as demonstrated previously and illustrated in Figures \ref{FigImagesArtefactsRADI}, \ref{FigImagesArtefactsRADICulminationEffect} and  \ref{FigImagesArtefactsRADIBreakEffect}. Therefore, any planet located in this region and that has a flux comparable to the disk flux will undergo severe flux losses from the disk itself in addition to its own self-subtraction. The position and amplitude of these negative areas are difficult to predict, therefore planet detection limits can be easily overestimated.  

\begin{figure}
   \centering
   \includegraphics[width=9cm]{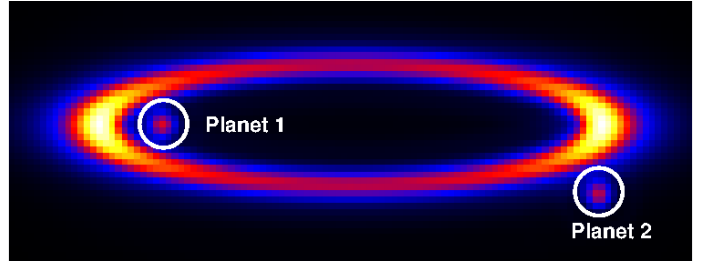}
   \caption{Position and brightness of the two fake planets relative to the disk model of HR\,4796A. The color scale is linear.}
         \label{FigPlanetPosition}
   \end{figure}

To give a quantitative example of this effect, we used the model of HR\,4796A disk and injected two fake faint planets at two critical locations where much flux loss from the disk self-subtraction is expected (see Figure \ref{FigPlanetPosition}). The planet fluxes were scaled relative to the disk brightness to correspond to a $1.4 M_{Jup}$ planet according to the evolutionary model of \citet{Baraffe2003}, or a flux ratio between the planet and the disk in the ansae of around 1:2. Then ADI reduction was performed in cADI and rADI for two amplitudes of field rotation $\Delta\theta=77.5^\circ$ and  $23^\circ$. The rADI parameters are a separation criterion $N_\delta=1.5$ FWHM, a number of images $N=20$ and an annulus width $dr=1.4$ FWHM. No noise was added to the data and we considered a baseline scenario of two planets alone without disk. We compare the results in terms of planet flux loss in Table \ref{TabPlanetFluxLoss}. The presence of the disk significantly impacts the detectability of both planets because the reference frame captures much flux from the disk at the position of the two planets. Although the planet self-subtraction is always well below 10\% even for a small field rotation, the presence of the disk very often leads to a 100\% flux loss, meaning that the planet is no longer detectable. 

\begin{table*}[t!]
\caption{Flux loss from two fake planets after ADI reduction. The flux loss is computed by aperture photometry with a radius of $0.75 \times FWHM$. mcADI is a technique introduced and commented on Section \ref{SubsectionMasking}}
\label{TabPlanetFluxLoss}
\begin{center}
\begin{tabular}{ c  c | c c c c | c c c c}

& & \multicolumn{4}{|c|}{$\Delta \theta = 77.5^\circ$} & \multicolumn{4}{c}{$\Delta \theta = 23^\circ$} \\
& & cADI & mcADI & rADI $N_\delta = 1.5$ & rADI $N_\delta = 3$ & cADI & mcADI & rADI $N_\delta = 1.5$ & rADI $N_\delta = 3$ \\
\hline
\multirow{2}{*}{Planet 1} & without disk & $0.2\%$ & $0.3\%$ & $1.4\%$ & $0.2\%$ & $8.5\%$ & NA & $2.7\%$ & $2.6\%$ \\
& with disk & $27.1\%$ & $4.3\%$ & $100\%$ & $23.5\%$ & $100\%$ & NA & $100\%$ & $99.7\%$ \\
\hline
\multirow{2}{*}{Planet 2} & without disk & $0.1\%$ & $0.3\%$ &$1.4\%$ & $1.2\%$ & $3.1\%$ & $12.9\%$ & $2.7\%$ & $0.7\%$ \\
& with disk & $4.4\%$ & $1.8\%$ & $100\%$ & $100\%$ & $63.0\%$ & $21.1\%$ & $45.0\%$ & $100\%$ \\

 \end{tabular}
\end{center}
\end{table*}

\section{Improvements to ADI methods: masking, iterating, and damping techniques}
\label{SectionImprovements}

Several ADI techniques more adapted for extended geometries are reviewed in this section. The biases they involve on disk observables are reported in Table \ref{TabSummary} and are commented on in the following subsections. 

\begin{table*}[t!]
\caption{Summary of the disk biases with standard and disk-specific ADI reduction strategies. Flux losses and FWHM changes were measured on data with and without speckle noise respectively, for an initial disk parametrized with $i=75^\circ$, $\alpha_{in}=5$ and $\alpha_{out}=-2.5$. The rADI and LOCI parameters are $N_\delta=1.5$, $N_A=300$, $g=1$, except for conservative LOCI, which has $N_\delta=2$, $N_A=1000$ and $g=0.5$}
\label{TabSummary}
\begin{center}
\begin{tabular}{ >{\centering}m{1.8cm} | >{\centering}m{1cm} | >{\centering}m{1cm} | >{\centering}m{1cm} | >{\centering}m{1cm} | >{\centering}m{1cm} | >{\centering}m{1cm} | >{\centering}m{1cm} | >{\centering}m{1cm} | >{\centering}m{1cm} | >{\centering}m{1cm} | >{\centering}m{1cm} }

& cADI & mcADI & cADI iter. 1 & cADI iter. 2 & rADI & LOCI & mLOCI & cons. LOCI & LOCI nnls & dLOCI $\Lambda=1$ & dLOCI $\Lambda=10$ \tabularnewline
\hline
Flux loss (\%) & 60.7 & 21.6 & 53.4 & 51.6 & 69.1 & 71.8 & 56.3 & 53.8 & 66.7 & 64.3 & 49.0 \tabularnewline
\hline
FWHM change (\%) & -21.7 & -3.3 & -0.3 & -2.4 & -32.7 & -24.9 & -21.9 & -13.6 & -31.4 & -29.5  & -21.2 \tabularnewline
 \end{tabular}
\end{center}
\end{table*}

\subsection{Iterations}

The main challenge of applying ADI on disks is to separate the disk from the star to build reference frames. The very first solution proposed to tackle this problem is an iterating process, applied to reveal the $\beta$ Pictoris disk from HST/STIS images \citep{Heap2000}. It was reapplied more recently for the disk of HD\,207129 on HST/ACS images \citep{Krist2010}. For those two space-based data sets, only two images at two different roll angles were used. However, one can apply this principle for a sequence of ground-based images taken in pupil-tracking mode. It is described for a single iteration in \citet{Lagrange2012}. A first estimate of the reference PSF image is made by median-combining all images and subtracting them from the images as in cADI. The residual image after derotation and collapse is the first estimate of the disk. In a second step, the contamination of the reference frame by the disk is computed using the estimated disk image of step 1. This contamination is then subtracted from the reference frame of step 1. PSF subtraction, derotation and collapse then yield the residual disk image of step 2. This process is then repeated iteratively. This equals applying sequentially cADI after correcting the reference frame from the residual image of the previous iteration.

Simulations on fake disks show that this process helps indeed to recover part of the disk flux lost by the effect of self-subtraction. More importantly, it can reduce the amplitude of the negative areas in the reduced image if one uses the a priori information that the estimate of the disk image must be positive. However, solutions converge toward a final disk image that is still less extended than the initial disk since information is lost in the areas where negative flux appears after the first guess.  
This process was applied to two real astrophysics ground-based observations: $\beta$ Pictoris \citep{Lagrange2012} and HR\,4796A \citep{Lagrange2012b}. Because the amplitude of field rotation was large, the gain was not significant and convergence was already reached after two iterations.   

It was described here for cADI, but it can also be used with rADI, the difference being that there is one reference frame for each image of the data cube. 

The typical flux recovery that can be expected for a disk inclined at $75^\circ$ with $\alpha_{in}=-2.5$ and $\alpha_{out}=-5$ with one and two iterations are given in Table \ref{TabSummary}. It shows that around 10\% more flux is preserved after the first two iterations. Close to the star, we emphasize, however, that it also amplifies residual structures that are not part of the disk, such as bright speckles or residuals of Airy rings. Far from the star, this techniques turns out to be efficient to preserve the initial disk surface brightness, since the measured FWHM along the semi-major axis is only slightly affected with changes smaller than 3\%.

\subsection{Masking the disk}
\label{SubsectionMasking} 
One of the drawbacks of iterating techniques is that if a bright speckle is present in the reduced disk image after the first iteration, it will be attributed to the disk extended emission. Therefore, it will stay there in the following iterations. One way to circumvent this difficulty is the following: instead of using the first iteration result as an a priori of the disk emission to subtract its contribution in the reference frame, one can also build an a priori binary mask of the disk and take into account only pixels outside this mask when building the reference frame. This obviously reduces the S/N, especially for very close regions, but the gain in outer regions, e.g. in the ring ansae, can be significant and negative areas are also limited. An example of masking in cADI is shown for HR\,4796A in \citet{Lagrange2012b}. It shows that the ansae are $10\%$ brighter with masking. 

Many aspects make this algorithm particularly attractive for disk reduction: 
\begin{enumerate}
\item The flux of the disk is well preserved for a well-defined mask (an additional 40\% of disk flux is recovered with respect to standard cADI, as shown in Table \ref{TabSummary}). Moreover, because negative regions are limited, the risk to overestimate planet detection limits is lower. This is shown in Table \ref{TabPlanetFluxLoss}: for a sufficient amplitude of field rotation, the flux loss is limited to $4.1\%$ for the closest planet (planet 1), whereas it reaches $27.1\%$ in cADI\footnote{We can also notice that flux losses are slightly higher in mcADI than in cADI without disk ($0.3\%$ vs $0.2\%$ for planet 1 and $0.3\%$ vs $0.1\%$ for planet 2). This is because when masking is applied, there are fewer frames available for the reference frame construction. Therefore, the relative number of frames with a small amplitude of field rotation is higher in mcADI than in cADI.}.   
\item The way the images are combined to build the reference frame is deterministic and independent of the noise distribution
\item All available images are used to build the reference frame, therefore fewer artifacts as in rADI are expected.
\item Using a disk binary mask is more relevant than using a separation criterion $N_\delta$ since this means adapting the criterion to the spatial extent of the disk. It requires that the disk is not too extended, however.
\end{enumerate} 

Masking the disk can also be applied with the LOCI reduction. Taking into account a binary weighting function in the LOCI algorithm is described in \citet{Pueyo2012} for point-source detection. It can be used for extended objects as long as the size of the optimization area after withdrawing masked pixels $N_A$ is still large enough for the inverse problem to be well-conditioned ($N_A \geqslant N_R$, where $N_R$ is the number of available reference images). As displayed in Table \ref{TabSummary}, the biases on disk observables have similar orders of magnitude as for a standard LOCI algorithm when conservative parameters are used ($N_\delta=2$, $N_A=1000$ and $g=0.5$).
   
\subsection{Damped LOCI}
\label{SubSectiondLOCI} 

Damped LOCI is an algorithm that forces the coefficients used to construct the reference frame to be positive (referred to as LOCI nnls in Table \ref{TabSummary} for non-negative least squares) and uses a modified cost function with respect to standard LOCI to compute them \citep{Pueyo2012}. It seeks to minimize the residual flux in the optimization region while maximizing the flux in the subtraction region. The maximization constraint is set by introducing an additional penalty term in the cost function. The relative importance of this term depends on the Lagrangian multiplier factor $\Lambda$. Therefore the final reduced image will be a trade-off between minimization of the speckle noise and preservation of the disk flux. It has only been applied to point-like sources so far and we propose to investigate its effect on extended objects. 

\begin{figure*}
   \centering
   \includegraphics[width=15cm]{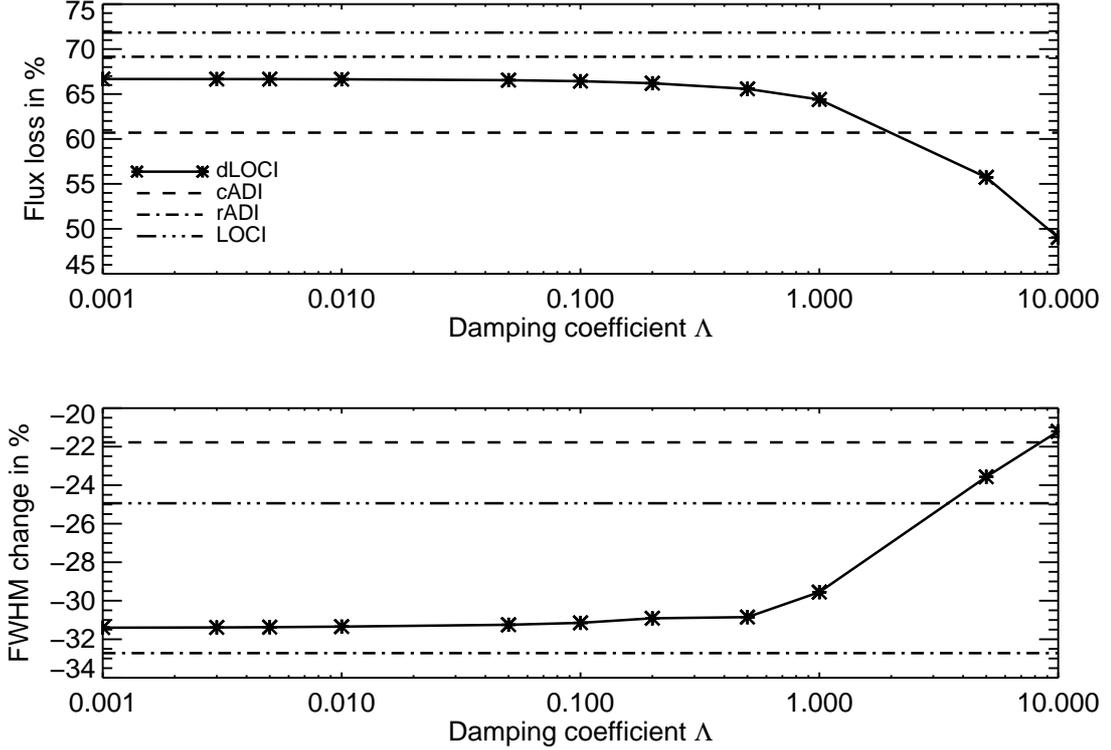}
   \caption{Impact of the damping coefficient $\Lambda$ on the  disk observables. The value of the parameters $N_\delta$, $g$ and $N_A$ are kept constant to $1.0$, $1.0$ and $300$, respectively. The flux change in cADI, rADI and standard LOCI is displayed for reference, the x-axis is on a logarithmic scale.}
         \label{FigChartInfluenceDampingCoeffTotalFlux}
   \end{figure*}

\begin{figure}
   \centering
   \includegraphics[width=9cm]{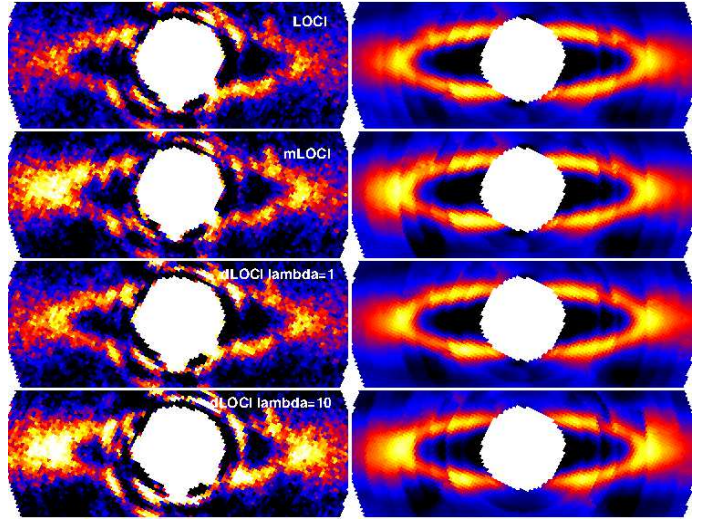}
   \caption{Right: comparison between standard, masked and damped LOCI ($\Lambda=1$ and $\Lambda=10$) for the same  $N_\delta=1.5$, $g=1$ and $N_A=300$ on a disk parametrized by $\alpha_{in}=5$, $\alpha_{out}=-2.5$ and $i=75^\circ$. Right: corresponding disks after applying the same LOCI coefficients to a data cube containing only the fake disk (without noise).}
         \label{FigdLOCIeffect}
   \end{figure}

Figure \ref{FigChartInfluenceDampingCoeffTotalFlux} shows that the constraint on the positivity of the LOCI coefficients has a strong effect on the preservation of the disk, with an 8\% gain on the recovered disk flux.  It also performs better than mLOCI. The impact of increasing the damping coefficient from 0 to 1 on the recovered disk flux is linear but relatively small (3\% additional gain). An additional increase in $\Lambda$ up to 10 enables one to gain even more flux at the expense of the speckle noise attenuation. On the measured FWHM, changes induced by dLOCI with $\Lambda \leq 3$ are slightly higher than for standard LOCI and reach the same level as cADI for $\Lambda=10$. A qualitative comparison of disks reduced with standard LOCI, mLOCI and dLOCI for $\Lambda=1$ and  $\Lambda=10$ is provided in Figure \ref{FigdLOCIeffect}. As explained, this shows that more disk flux is preserved in mLOCI and dLOCI.  However, we also clearly see that speckles pinned on the diffraction rings are less well subtracted in those cases. Bright blobs described in Section \ref{Sect:cADIeffects} appear in the two ansae for high damping coefficient (dLOCI, $\Lambda=10$) or in mLOCI.

\section{Conclusions}

We discussed the effects of ADI applied to extended circumstellar disks. A theoretical study demonstrated that the field of application of this imaging technique depends on two parameters: the available amplitude of field rotation and the geometrical extension of the object in azimuth. This analysis gives a clear indication of the required amplitude of field rotation to reduce a given disk depending on its morphology, and conversely it gives an indication of the disk minimum inclination for a given field rotation.
With reasonable observing time, objects inclined less than $50^\circ$ are out of the scope of ADI. For all others, self-subtraction is a problem that observers must bear in mind when reducing the data. It affects the whole geometry of the disk by reducing the total disk flux, but also introduces local effects difficult to distinguish from speckle noise patterns. A typical effect already observed for some circumstellar disks reduced with ADI is the enhancement of the ansae for an inclined ring morphology, creating bright blobs along the disk major axis that can be easily misinterpreted. Brightness asymmetries from non-uniform parallactic angle evolution are other striking effects that arise from ADI, especially rADI in which a limited number of images is used to build the reference frames. 

Measurable astrophysical quantities used to characterize disks such as inner/outer slopes of the SBD or ring FWHM are biased and their estimation must be calibrated. We here provided an insight into how these parameters are affected, depending on the initial disk morphology and the reduction parameters.

To evaluate the efficiency of the explored ADI algorithms applied on disks, two key points must be considered: the amount of disk self-subtraction and how well the residual speckles are attenuated. For the first point, masking as used in mcADI is our preferred data reduction strategy, since it replaces the classical separation criterion $N_\delta$ appropriate for point sources with a binary mask adapted to the expected extension of the object, to efficiently limit self-subtraction. The second point is better addressed with optimization algorithms such as LOCI, but flux losses are much more severe. Our study shows that using larger optimization areas, with an azimuthal extent greater than the radial extent, helps to preserve the disk flux.  More importantly constraining the LOCI coefficients to be positive proves to be very effective in this goal. 

\begin{acknowledgements}
      We would like to thank J-C Augereau for letting us use the GRaTer code to simulate fake disk images.
\end{acknowledgements}

\bibliography{Milli2012} 
\Online
\begin{appendix} 
\section{Deriving the flux loss criteria in cADI}
\label{AppendixFluxLossCriteria} 

\begin{figure}
	\centering
	\includegraphics[width=9cm]{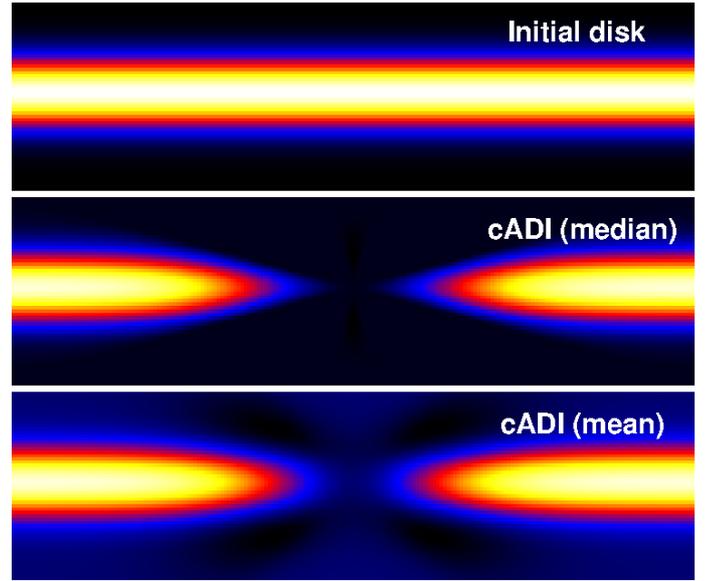}
	\caption{Theoretical edge-on disk.}
	\label{FigDiskCadiMedCadiMean}
\end{figure}

We derive the expression of the residual flux given in Equation \ref{EquFluxLoss}, which is valid for a disk seen edge-on, with an intensity vertical profile parametrized by the expression  $I(z)=e^{-{\left( \frac{z}{h} \right)} ^\gamma}$. Such a disk is plotted in Figure \ref{FigDiskCadiMedCadiMean}. The vertical scale height $h$ is in this case, constant with the radius, but the following derivation still holds if h varies with the separation to the star, for example if there is a flaring characterized by a flaring index $\beta$ such that $h=h(x)=\left(\frac{x}{x_0}\right)^\beta$. Similarly, the intensity in the disk mid-plane ($z=0$) is constant here and equal to 1, but the derivation still holds if the intensity  varies with the separation $x$.

\begin{figure*}
\centering
\includegraphics[width=15cm]{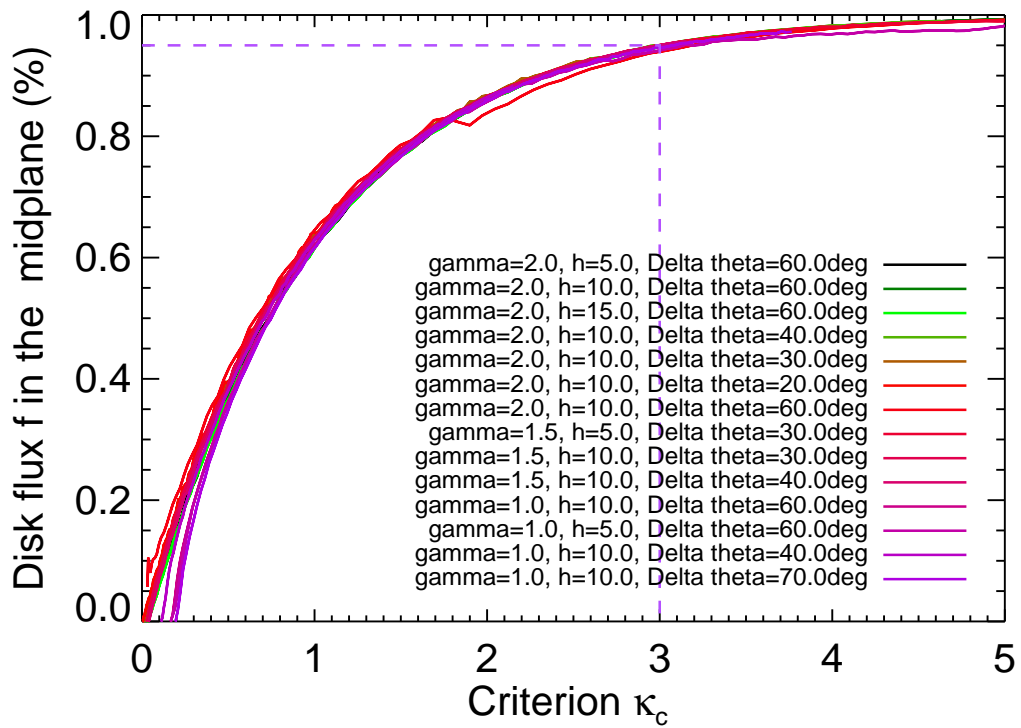}
\caption{Experimental verification of Equation \ref{EquFluxLoss} on simulated disks (with different values of the vertical exponent $\gamma$ and the vertical height h) for various field rotations $\Delta \theta$. For $\kappa_c=3$, $95\%$ of the disk flux f is preserved.}
\label{FigFluxLossVsKappaCadi}
\end{figure*}

We consider a point A located at a separation $x$ and a height $z$. It makes an angle $i=arctan \left( \frac{z}{x} \right)$ with the disk mid-plane. 
The vertical profile at this separation as a function of i is displayed in Figure \ref{FigFluxLossVsKappaCadi}. The two blue arrows represent the amplitude of field rotation $\Delta \theta/2$ on each side of A. In cADI, the flux of the reference frame at the position of A is equal to the median of the curve between  $i-\Delta \theta/2$ and  $i+\Delta \theta/2$, sampled according to the parallactic angle distribution. We assumed that this sampling is uniformly distributed. To compute the median, we divide this range into four regions, according to the values of the intensity:
\begin{itemize}
\item region 1, of length $2i$, the farthest on the right characterized by the lowest intensity values
\item regions 2 and 3, symmetrical about the mid-plane, of length $\Delta \theta/2-2i$
\item  region 4, of length $2i$ centered on the mid-plane and consisting of the highest intensity values.
\end{itemize}

This decomposition into four regions of increasing intensity allows us to directly conclude that the median over the four regions is also equal to the median over region 2 only, which is equal to the intensity at point B for $\Delta \theta/4$. 

The disk flux $f$ after subtraction of the reference frame can then be expressed as
\begin{eqnarray*}
f  	&=& \frac{I(r,i)-I(r,\Delta \theta/4)}{I(r,i)} \\
   & = & \frac{exp \left(-(\frac{i}{\Delta \alpha})^\gamma \right) - exp \left(-(\frac{\Delta \theta/4}{\Delta \alpha})^\gamma \right)}{exp \left(-(\frac{i}{\Delta \alpha})^\gamma \right)} \\
   & =& 1 - exp \left(-\frac{(\Delta \theta/4)^\gamma - i^\gamma}{\Delta \alpha^\gamma} \right) \\
   & =& 1- e^{-\kappa_c}.\\
\end{eqnarray*}

   \begin{figure*}
   \centering
\includegraphics[width=15cm]{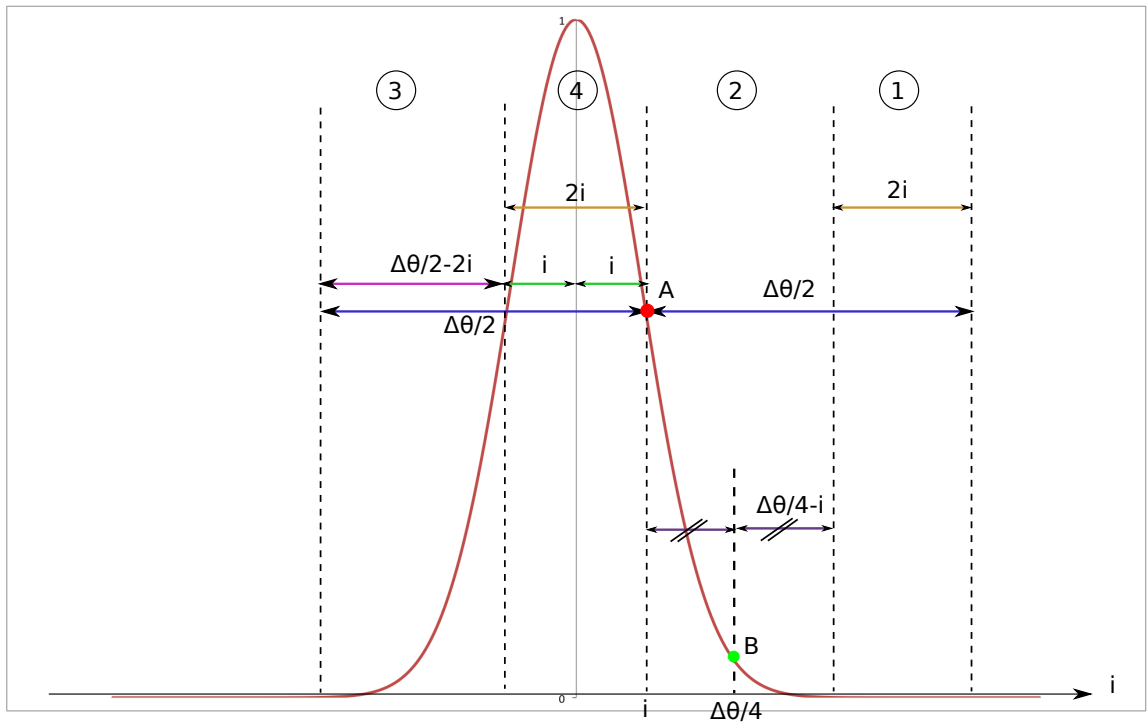} 
\caption{Intensity vertical profile against the angular height i.}
\label{FigGaussian}
\end{figure*}

An experimental verification of this law was performed and is shown in Figure \ref{FigFluxLossVsKappaCadi}. The disk flux $f$ in the mid-plane ($i=0$) after cADI reduction, expressed in percentage for various disks and field rotations, is plotted against the criterion $\kappa_c$. All curves match well, which confirms the derived theoretical expression. A similar graph can be made for different vertical profiles instead of mid-plane profiles. 

\end{appendix}

\end{document}